\documentclass[12pt,english]{article}
\usepackage[T1]{fontenc}
\usepackage[latin9]{inputenc}
\usepackage[a4paper,textwidth=16.04cm,textheight=22cm,footskip=15mm]{geometry}
\usepackage{amsfonts}
\usepackage{graphicx}
\usepackage{xcolor}
\usepackage{mathtools}
\usepackage{bm}
\usepackage{dsfont}
\usepackage[hidelinks]{hyperref}
\usepackage{cite} 

\makeatletter

\makeatletter

\newcommand{\lyxaddress}[1]{
\par {\raggedright #1
\vspace{1.4em}
\noindent\par}
}

\date{}

\makeatother

\usepackage{babel}

\begin{document}

\title{
\vspace{2cm}
A note on scaling arguments in the effective average action formalism}

\vspace{1cm}

\author{Carlo Pagani}

\maketitle

\lyxaddress{\begin{center}
Institut f\"{u}r Physik, PRISMA \& MITP Johannes-Gutenberg-Universit\"{a}t, \hspace{2cm}
Staudingerweg 7, 55099 Mainz, Germany
\par\end{center}}


\begin{abstract}
The effective average action (EAA) is a scale dependent effective
action where a scale $k$ is introduced via an infrared regulator.
The $k-$dependence of the EAA is governed by an exact flow equation to which one associates a boundary condition at a scale $\mu$. 
We show that the $\mu-$dependence of the EAA is controlled by an equation fully analogous to the Callan-Symanzik equation which allows to define scaling quantities straightforwardly. 
Particular attention is paid to composite operators which are introduced along
with new sources. We discuss some simple solutions to the flow equation
for composite operators and comment their implications in the case
of a local potential approximation. 
\end{abstract}


\section{Introduction}

The renormalization group (RG) provides an ideal framework to discuss
the scaling of operators in quantum field theory. In this work we consider
the scaling properties of a quantum field theory within
the effective average action (EAA) formalism, which is a functional
realization of the Wilsonian renormalization program \cite{Wetterich:1992yh,Ellwanger:1993mw,Morris:1993qb}. 
The EAA is a scale dependent effective action whose associated action has been
modified by the introduction of an infrared cutoff depending on a
scale $k$ in such a way that low momentum modes are suppressed in the integration. 
In order to discuss the scaling properties within this formalism
we show that the EAA satisfies, besides an exact flow equation involving the
scale $k$, an equation which involves the scale $\mu$ at which the
boundary conditions are imposed. This equation entails an invariance
under changes of ``floating normalization point'' $\mu$ very
similar to one expressed by the Callan-Symanzik equation. 
In this sense this equation recalls the connection between the methods
of perturbative renormalization and the Wilsonian RG studied in \cite{Benettin:1976qz}.

We carefully investigate composite operator by modifying the original
action with a source dependent term which allows us to consider composite
operators insertions via functional derivatives with respect to the
source. We will see that the above mentioned $\mu-$invariance allows defining scaling
operators straightforwardly. Then we comment the relation between
the so defined scaling dimension and critical exponents and show that,
if total derivative terms are neglected, the results are easily related.
Moreover we will show that introducing the sources for composite operators
allows us to identify some general types of composite operators among
which the descendant operators of the scaling operators at the fixed
point. To make our discussion concrete we revisit the result of the
local potential approximation under our perspective. 

The paper is organized as follows. 
In section \ref{sec:RG-flow-of-composite-operators}
we introduce the flow equation for the EAA and for composite operators.
In section \ref{sec:Floating-normalization-point-and-scaling} we
discuss the $\mu-$(in)dependence of the EAA and how scaling dimensions
can be identified while in section \ref{sec:Scaling-solutions-and-composite-operators}
we consider the local potential approximation in view of the previous
discussion. In section \ref{sec:Conclusions-and-outlooks} we summarize
our results and discuss possible outlooks.

\section{RG flow of composite operators \label{sec:RG-flow-of-composite-operators}}

In quantum field theory a composite operator is defined as a (local)
function of the field and its derivatives, i.e.~$O=O\left(\varphi,\partial_{\mu}\varphi\right)$.
Generically, once such an operator is inserted in a correlation function
new divergences appear. Owing to these divergences one has to renormalize,
besides the couplings of the theory, also the composite operator itself.
As a result one finds that a renormalized composite operator, which we denote $\left[O\right]$
using square brackets, is a sum of various operators. For instance,
at one loop in a six dimensional $\varphi^{3}$ theory one finds $\left[\varphi^{2}\right]=c_{1}\varphi +c_{2}\Delta\varphi+c_{3}\varphi^{2}$,
where $c_{i}$ are suitable coefficients \cite{Collins:1984xc}. A convenient
way to keep track of composite operators and their insertion into
Green's functions is to couple them to a source $\varepsilon$ adding
a term $\varepsilon\cdot O$ to the action as follows\footnote{Whenever a dot appears in a mathematical expression, e.g.: $f\cdot g$,
the DeWitt notation is used meaning that integration and index summation
is intended.}: 
\begin{eqnarray*}
\langle O\rangle & = & {\cal N} \int{\cal D}\chi\,Oe^{-S}\\
 & = & -\frac{\delta}{\delta\varepsilon_{x}} {\cal N} \int{\cal D}\chi\,e^{-S-\varepsilon\cdot O}\, ,
\end{eqnarray*}
where ${\cal N}$ is a suitable normalization.
Let us consider the generating functional $W\left[J,\varepsilon\right]$
for the connected Green's functions associated to the modified action $S+\varepsilon \cdot O$:
\begin{eqnarray*}
e^{W\left[J,\varepsilon\right]} & \equiv & \int{\cal D}\chi\,e^{-S-\varepsilon\cdot O+J\cdot\chi}\,.
\end{eqnarray*}
The connected part of the correlation function $\langle O\rangle$
will be given just by the derivative $-\frac{\delta}{\delta\varepsilon}W\left[J,\varepsilon\right]$.
Via a Legendre transform we obtain the associated effective action
(EA), $\Gamma\left[\varphi,\varepsilon\right]$:
\begin{eqnarray*}
\Gamma\left[\varphi,\varepsilon\right] & = & J\cdot\varphi-W\left[J,\varepsilon\right]\,,\;\varphi=\delta_{J}W\,.
\end{eqnarray*}
Note that we do not perform a Legendre transform with respect to the
source $\varepsilon$.\footnote{
Such a Legendre transform is performed on a bilocal source when
considering the 2PI effective action \cite{Cornwall:1974vz} after adding a term $\varphi (x) \cdot \varepsilon (x,y) \cdot \varphi (y)$ to the action.} 
Since the insertion
of one composite operator is related to a functional differentiation
with respect to the source $\varepsilon$ let us consider: 
\begin{eqnarray}
\delta_{\varepsilon}\Gamma\left[\varphi,\varepsilon\right] & = & \delta_{\varepsilon}\left(J\cdot\varphi-W\left[J,\varepsilon\right]\right)\nonumber \\
 & = & \frac{\delta J}{\delta\varepsilon}\cdot\varphi-\left(\frac{\delta W}{\delta\varepsilon}\left[J,\varepsilon\right]+\frac{\delta J}{\delta\varepsilon} \cdot \delta_{J}W\right)\nonumber \\
\frac{\delta\Gamma}{\delta\varepsilon}\left[\varphi,\varepsilon\right] & = & -\frac{\delta W}{\delta\varepsilon}\left[J,\varepsilon\right]\,.\label{eq:relation_Gamma_W}
\end{eqnarray}
This tells us that we can obtain information regarding the renormalization
of composite operators by considering the renormalization of $\delta_{\varepsilon}\Gamma\left[\varphi,\varepsilon\right]$.

As already said we shall work within the functional renormalization
group (FRG) framework. In particular we consider the RG flow of the effective
average action (EAA) which is a scale dependent generalization of
the standard effective action. One first defines a modified generating
functional of connected Green\textquoteright s functions,  $W_{k}\left[J\right]$:
\begin{eqnarray*}
e^{W_{k}\left[J\right]} & \equiv & \int{\cal D}\chi\,e^{-S-\Delta S_{k}+J\cdot\chi} \, ,
\end{eqnarray*}
where $\Delta S_{k}$ suppresses the integration of momentum modes
$p^{2}<k^{2}$ and is quadratic in the fields with a kernel $R_{k}$,
i.e.~$\Delta S_{k}=\frac{1}{2}\int\chi R_{k}\chi$. Note that this
cutoff action acts like an infrared cutoff. Let us denote $\tilde{\Gamma}_{k}$
the Legendre transform of $W_{k}$ and define the EAA subtracting
the cutoff action which we added at the beginning: 
\begin{eqnarray*}
\Gamma_{k} & \equiv & \tilde{\Gamma}_{k}-\Delta S_{k}\,.
\end{eqnarray*}
The $k-$dependence of the functional $\Gamma_{k}$ satisfies the following
exact equation \cite{Wetterich:1992yh,Ellwanger:1993mw,Morris:1993qb}: 
\begin{eqnarray}
\partial_{t}\Gamma_{k} & = & \frac{1}{2}\mbox{Tr}\left[\left(\Gamma_{k}^{\left(2\right)}+R_{k}\right)^{-1}\partial_{t}R_{k}\right]\,,\label{eq:flow_eq_EAA}
\end{eqnarray}
where $\partial_{t}=k\partial_{k}$ is the logarithmic derivative
with respect to the cutoff and $\Gamma_{k}^{\left(2\right)}$ is the
Hessian of the EAA. To concretely employ equation (\ref{eq:flow_eq_EAA})
one needs to resort to some approximations and makes an ansatz for
$\Gamma_{k}$. Such procedure has been proved robust in many fields,
especially to determine scaling properties of statistical systems at criticality,
see \cite{Berges:2000ew,Delamotte:2007pf} for an overview.

The inclusion of composite operators in this framework is straightforward:
we simply upgrade the above definitions employing the modified action
$S+\varepsilon\cdot O$ instead of $S$ (we refer to \cite{Igarashi:2009tj} 
for a detailed discussion regarding the flow of composite operators, see also \cite{Pawlowski:2005xe,Rosten:2010vm}).
In this manner we obtain a modified generating functional
$W_{k}\left[J,\varepsilon\right]$ for connected Green's functions which depends also on the source
$\varepsilon$. In full analogy with the derivation of relation (\ref{eq:relation_Gamma_W})
we obtain:
\begin{eqnarray*}
 -\frac{\delta}{\delta\varepsilon}W_{k}\left[J,\varepsilon\right] & = & \frac{\delta}{\delta\varepsilon}\Gamma_{k}\left[\varphi,\varepsilon\right]\,.
\end{eqnarray*}
We also introduce the notation
\begin{eqnarray*}
\left[O_k \right] & \equiv & \frac{\delta}{\delta\varepsilon}\Gamma_{k}\left[\varphi,\varepsilon\right]\, ,
\end{eqnarray*}
where the subscript $k$ indicates that $\left[O_k \right]$ depends on the scale $k$.
In order to obtain the scale dependence of composite operators
we observe that the flow equation (\ref{eq:flow_eq_EAA}) holds in
full generality also for the modified EAA $\Gamma_{k}\left[\varphi,\varepsilon\right]$:
\begin{eqnarray}
\partial_{t}\Gamma_{k}\left[\varphi,\varepsilon\right] & = & \frac{1}{2} \mbox{Tr}
\left[\left(\Gamma_{k}^{\left(2\right)}\left[\varphi,\varepsilon\right]+R_{k}\right)^{-1}\partial_{t}R_{k}\right]\label{eq:flow_eq_EAA_epsilon}
\end{eqnarray}
where $\Gamma_{k}^{\left(2\right)}\left[\varphi,\varepsilon\right]$
denotes the Hessian of the EAA with respect to the field $\varphi$.
In order to obtain the renormalization regarding the insertion of
a single composite operator we have to consider a functional derivative
with respect to the source $\varepsilon$ and set $\varepsilon=0$
afterwards. In this way we obtain:
\begin{eqnarray*}
\partial_{t}\left(\frac{\delta}{\delta\varepsilon}\Gamma_{k}\left[\varphi,\varepsilon\right]\right)\Bigr|_{\varepsilon=0} & = & -\frac{1}{2} \mbox{Tr}
\left[\left(\Gamma_{k}^{\left(2\right)}+R_{k}\right)^{-1}\frac{\delta\Gamma_{k}^{\left(2\right)}}{\delta\varepsilon}\left(\Gamma_{k}^{\left(2\right)}+R_{k}\right)^{-1}\partial_{t}R_{k}\right]\Bigr|_{\varepsilon=0}\,.
\end{eqnarray*}
Let us observe that we can avoid performing a functional derivative
with respect to the source $\varepsilon$ and just compare order by
order in $\varepsilon$. Clearly, since we are interested just in
one insertion of the composite operator, we can limit ourselves to
consider an $\varepsilon$ such that $\varepsilon^{2}=0$ (the 
flow equation of order $O\left(\varepsilon^{2}\right)$ is considered in appendix \ref{sec:eps2-terms}). 
We can rewrite the flow equation as:
\begin{eqnarray}
\partial_{t}\left(\varepsilon\cdot \left[ O_{k} \right]\right) & = & -\frac{1}{2} \mbox{Tr}
\left[\left(\Gamma_{k}^{\left(2\right)}+R_{k}\right)^{-1}\left(\varepsilon\cdot \left[ O_{k}\right]^{\left(2\right)}\right)\left(\Gamma_{k}^{\left(2\right)}+R_{k}\right)^{-1}\partial_{t}R_{k}\right]\,.\label{eq:flow_eq_composite_operator_epsilon}
\end{eqnarray}
Equation (\ref{eq:flow_eq_composite_operator_epsilon}) can be seen
as an RG improved version of a one loop equation. To see this let
us consider the one loop EAA $\Gamma_{k,1}$ associated to the modified
action $S+\varepsilon\cdot O$: 
\begin{eqnarray*}
\Gamma_{k,1} & = & S+\varepsilon\cdot O+\frac{1}{2}\mbox{Tr}\log\left(S^{\left(2\right)}+R_{k}+\varepsilon\cdot O^{\left(2\right)}\right)  \, ,
\end{eqnarray*}
where $S^{\left(2\right)}$ and $O^{\left(2\right)}$ are the Hessians
for the action and the composite operator respectively. If we now
differentiate the above expression with respect to the scale $k$ we obtain:
\begin{eqnarray*}
\partial_{t}\Gamma_{k,1} & = & \frac{1}{2}\mbox{Tr}\left[\left(S^{\left(2\right)}+R_{k}+\varepsilon\cdot O^{\left(2\right)}\right)^{-1}\partial_{t}R_{k}\right]\,.
\end{eqnarray*}
Finally, in order to derive the running of the composite operator,
we just need to take a functional derivative with respect to the source
$\varepsilon$ and set this to zero: 
\begin{eqnarray}
\partial_{t} \left[O_k \right] & = & -\frac{1}{2}\mbox{Tr}\left[\left(S^{\left(2\right)}+R_{k}\right)^{-1}O^{\left(2\right)}\left(S^{\left(2\right)}+R_{k}\right)^{-1}\partial_{t}R_{k}\right]\,.\label{eq:flow_Composite_Operator_1_loop}
\end{eqnarray}
Equation (\ref{eq:flow_Composite_Operator_1_loop}) is fully analogous
to equation (\ref{eq:flow_eq_composite_operator_epsilon}) with the
microscopic action $S$ in place of the EAA and the bare operator
in place of the renormalized one.

As in the case of the flow equation (\ref{eq:flow_eq_EAA}), also the flow equation (\ref{eq:flow_eq_composite_operator_epsilon})
has to be equipped with suitable boundary conditions, see also \cite{Igarashi:2009tj}. 
When the scale $k$ has been lowered to zero
the integration has been fully performed and we have that $\langle O_{B}\rangle=\left[ O_{k=0} \right]$, 
where $O_B$ is the bare operator
(see \cite{Rosten:2006qx} for an analogous observation using the Wilsonian action). 

Finally we would like to stress the following point: in principle
the renormalization of composite operators must be carried out in
addition to the usual renormalization and some ``extra'' work is
needed. Since equation (\ref{eq:flow_eq_composite_operator_epsilon})
is essentially the linearization of the flow equation one may get the
impression that the operator dimensions at the fixed point are given
by the linearization of the RG, i.e.~by the critical exponents.\footnote{
Here critical exponents are associated to deformations of the type
$S+\delta S$ and not of the type $S+g\cdot\delta S$ where $g$ is
an arbitrary source. The latter case is considered by Wilson and Kogut
in \cite{Wilson:1973jj}. However in FRG computations one usually takes source
independent deformations.} 
Although critical exponents and scaling dimensions of composite operators are closely related
(as we will discuss in sections \ref{sub:Scaling-dimension-from-flow-eq}
and \ref{sub:Scaling-solutions}) there are some notable differences.
To understand why this is the case let us consider a simple example:
in a six dimensional $\varphi^{3}$ theory the operator $\left[\varphi^{2}\right]$
contains the operator $\Delta \varphi$ \cite{Collins:1984xc}, with $\Delta = -\partial^2$. If we consider
$\Delta\varphi$ as a term appearing in the action, this would result
simply in a surface term and as such this term would usually be neglected.
When considering composite operators it is no longer so, given that
$\varepsilon\cdot\Delta\varphi$ is not a surface term. For the same
reason one should distinguish between $\varphi\Delta\varphi$ and $\partial\varphi\partial\varphi$
when dealing with composite operators, in contrast to what one does
with couplings appearing in the action.
In the standard formulation of quantum field theory one indeed considers \cite{Itzykson:1980rh}:
\begin{eqnarray}
\varepsilon\cdot\left[\varphi^{4}\right] & = & \varepsilon\cdot\left(Z_{21}Z_{\varphi}\varphi^{2}+Z_{22}Z_{\varphi}^{2}\varphi^{4}+Z_{23}Z_{\varphi}\left(\partial\varphi\right)^{2}+Z_{24}Z_{\varphi}\varphi\Delta\varphi\right) \, ,
\label{eq:composite_operator_example_ItkZub}
\end{eqnarray}
where $Z_{\varphi}$ is the wave function renormalization.

Furthermore, equation (\ref{eq:flow_eq_composite_operator_epsilon})
can be used not only for scalar operators, but also for higher spin
ones. In this case the source will carry an index, for instance $\varepsilon^{\mu}$
in the case of a spin one operator. It appears clear that the renormalization
of such a term cannot be extracted directly from the renormalization
of the theory alone and its linearization around the fixed point.
Finally it may happen that one is interested in the flow of some particular operator
while neglecting some others. In such case our framework is particularly convenient,
see \cite{Pagani:2015hna} for an example of such an application.

\section{Floating normalization point and scaling dimensions \label{sec:Floating-normalization-point-and-scaling}}

In this section we discuss how RG quantities and scaling dimensions are related
in the functional renormalization group framework. In particular we
shall show that it is possible to infer an equation which resembles
the Callan-Symanzik equation. 
By reformulating scaling arguments via the Callan-Symanzik equation
we will obtain a generic framework which can be applied to any system
of interest. Besides this, our equation shows an obvious connection
with the standard framework of quantum field theory and may be a useful tool for comparison
with results obtained via other approaches. 
In section \ref{sub:RG-invariance-of-fRG} we describe the invariance of
the EAA with respect to suitable changes of boundary conditions of
the flow and derive a Callan-Symanzik type of equation. In section
\ref{sub:Fixed-point-and-anomalous-dim} we extend those arguments
including composite operators, while in section \ref{sub:Scaling-dimension-from-flow-eq}
we relate these results to the usual quantities computed via the
FRG.

We will often work within the so called LPA$^{\prime}$ truncation
where one takes into account up to two derivatives of the field and
a generic potential including the wave function renormalization of
the field:
\begin{eqnarray}
\Gamma_{k}\left[\varphi\right] & = & \int\left[\frac{Z_{k}}{2}\partial_{\mu}\varphi\partial^{\mu}\varphi+U_{k}\left(Z_{k}^{1/2}\varphi\right)\right]\,.\label{eq:LPAprime_ansatz_Zevery_field}
\end{eqnarray}
The extension of our arguments to more general truncations is obvious.
In the above ansatz we made explicit the fact that the wave function
renormalization can be looked at as an inessential coupling and can
be removed via a rescaling of the field (see \cite{Rosten:2010vm} for a
similar discussion in the context of the Wilsonian action).
One can eventually define the field $\phi=Z_{k}^{1/2}\varphi$ and
insist on having a canonically normalized kinetic term. When the flow
equation is expressed in terms of $\phi$ the effect of the wave function
renormalization is fully contained in the appearance of the anomalous
dimension. In this work we shall express the EAA as a function of
$\varphi$ or $\phi$ according to convenience.

\subsection{Invariance under changes of the floating normalization point \label{sub:RG-invariance-of-fRG}}

\begin{figure}[h]
\begin{centering}
\includegraphics[scale=0.25]{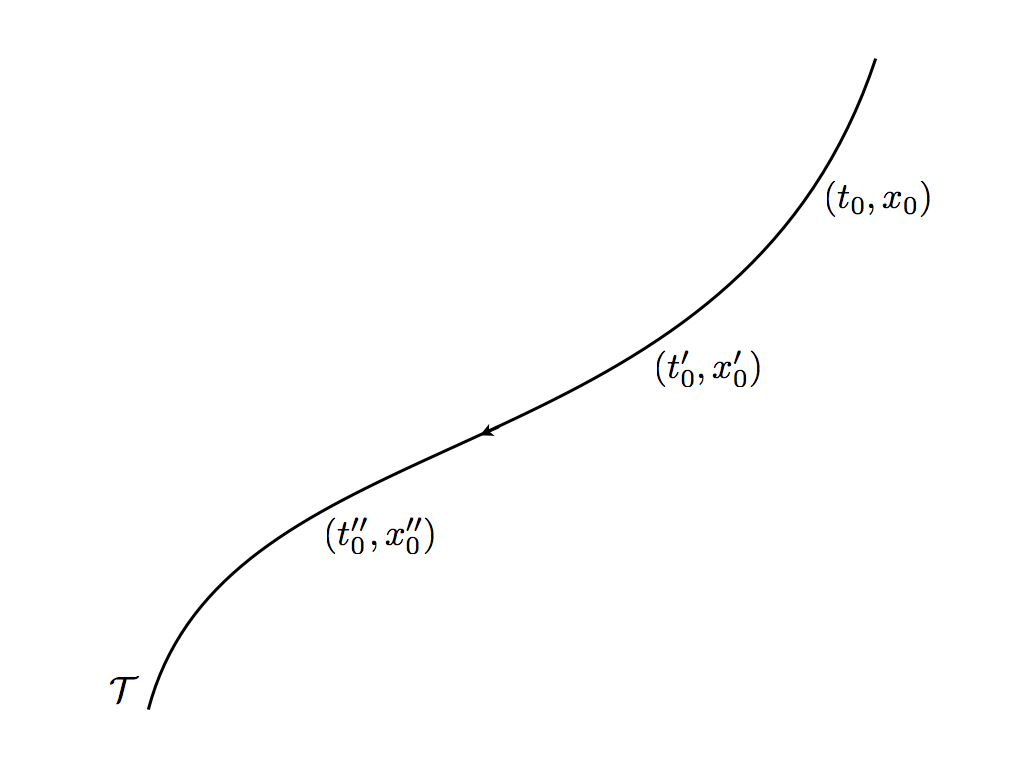} 
\par
\end{centering}
\caption{Trajectory of the dynamical system (\ref{eq:generic_dyn_syst}). The couples $(t_0,x_0)$, $(t_0^{\prime},x_0^{\prime})$ and
$(t_0^{\prime\prime},x_0^{\prime\prime})$ are possible boundary conditions associated to the trajectory $\cal{T}$.}
\label{img:sist_din_traj}
\end{figure}

For the sake of simplicity let $U_{k}$ be a polynomial of a given
order whose coefficients are parametrized by a set of dimensionful
couplings $\left\{ g_{i}\right\} $, whose dimensionless version is
denoted $\left\{ \tilde{g}_{i}\right\} $. The RG flow is given by
a set of first order differential equations of the type
\begin{eqnarray*}
\partial_{t}\tilde{g}_{i} & = & f_{i}\left(\left\{ \tilde{g}_{j}\right\} \right)\,.
\end{eqnarray*}
Let us consider a given trajectory which is associated to a boundary
condition 
\begin{eqnarray*}
\tilde{g}_{i}\left(\mu\right) & = & \tilde{g}^{\left(R\right)}
\end{eqnarray*}
at the scale $\mu$, which we could call ``floating normalization point''.
We observe that this trajectory can be labelled by many other equivalent
boundary conditions along the RG trajectory at some other scale $\mu^{\prime}$.
In order to make this clear let us consider a simple example of a
dynamical system which mimics our system of beta functions:
\begin{eqnarray}
 & \left\{ \begin{array}{c}
\dot{x}\left(t\right)=f\left(x\right)\\
x\left(t_{0}\right)=x_{0}\,\,\,\quad
\end{array}\right.\,. \label{eq:generic_dyn_syst}
\end{eqnarray}
Let $F\left(t\right)$ be the generic solution of the upper equation in (\ref{eq:generic_dyn_syst})
and suppose that we can implement the boundary condition $x\left(t_{0}\right)=x_{0}$
explicitly via:
\begin{eqnarray}
x\left(t\right) & = & F\left(t\right)-F\left(t_{0}\right)+x_{0}\,.  \label{eq:sol-of-simple-dyn-syst}
\end{eqnarray}
The fact that we can associate to this trajectory many
other boundary conditions is expressed by the vanishing of the total
derivative with respect to $t_{0}$:
\begin{eqnarray*}
\frac{d}{dt_{0}}x\left(t\right) & = & \frac{d}{dt_{0}}\left[F\left(t\right)-F\left(t_{0}\right)+x_{0}\right]\\
 & = & -\frac{d}{dt_{0}}F\left(t_{0}\right)+\dot{x}_{0}\\
 & = & -\dot{x}_{0}+\dot{x}_{0}=0\,.
\end{eqnarray*}
In the second line we took care of both the explicit presence of $t_{0}$
in the first term and of the implicit dependence of $x_{0}$ on the
``normalization point'' $t_{0}$. Thus we can rewrite the above equation
as:
\begin{eqnarray}
\frac{d}{dt_{0}}x\left(t\right) & = & \left(\frac{\partial}{\partial t_{0}}+\dot{x}_{0}\frac{\partial}{\partial x_{0}}\right)x\left(t\right)=0\,.\label{eq:Invariance-under-BC-in-SistDin}
\end{eqnarray}
Note that the boundary condition $x_{0}$ may enter in the solution
in many possible ways and not just as shown in (\ref{eq:sol-of-simple-dyn-syst}). 
The invariance under the operator in the RHS
of equation (\ref{eq:Invariance-under-BC-in-SistDin}) is guaranteed
by the following reasoning: given a trajectory (i.e.~a solution of
the equation) associated to the boundary condition $x_{0}$ at $t_{0}$, we observe
that the same trajectory can be associated to many other boundary
conditions along the same trajectory at some other points $x_{0}^{\prime},x_{0}^{\prime\prime},\cdots$,
see figure \ref{img:sist_din_traj}. 
This means that the selected trajectory $x\left(t\right)$ is invariant
under translations of the boundary condition along the trajectory itself. Thus the
solution is not invariant under any change of the couple $\left(t_{0},x_{0}\right)$
but is invariant under those changes which bring $\left(t_{0},x_{0}\right)$
into another point along the solution. Indeed this is exactly what
is implemented by the operator of equation (\ref{eq:Invariance-under-BC-in-SistDin}).
To see this let us make explicit the presence of $\left(t_{0},x_{0}\right)$
in the solution denoting $x\left(t\right)=x\left(t;t_{0},x_{0}\right)$.
Let us translate $t_{0}$ infinitesimally and move $x_{0}$ accordingly:
\begin{eqnarray*}
x\left(t;t_{0},x_{0}\right) & = & x\left(t;t_{0}+\varepsilon,x_{0}+\delta x_{0}\right)\\
 & = & x\left(t;t_{0}+\varepsilon,x_{0}+\dot{x}_{0}\varepsilon\right)\\
 & = & x\left(t;t_{0},x_{0}\right)+\varepsilon\left(\partial_{t_{0}}+\dot{x}_{0}\partial_{x_{0}}\right)x\left(t;t_{0},x_{0}\right) \, ,
\end{eqnarray*}
where we used $x_{0}=x\left(t_{0}\right)$ and so $\delta x_{0}=\dot{x}\left(t_{0}\right)\varepsilon$.
We can rewrite the above equation as
\begin{eqnarray}
\left(\partial_{t_{0}}+f\left(x_{0}\right)\partial_{x_{0}}\right)x\left(t\right) & = & 0  \, , \label{eq:CS-for-dyn-syst}
\end{eqnarray}
where we exploited $\dot{x}_{0}=f\left(x_{0}\right)$.

There is nothing preventing us from applying this reasoning directly to
our system of equation describing the RG flow of the EAA. Let us denote
with the superscript $\left(R\right)$ all the boundary conditions,
for example $\tilde{g}^{\left(R\right)}=\tilde{g}\left(\mu\right)$
is the boundary condition associated to $\tilde{g}\left(k\right)$.
The analogue of equation (\ref{eq:CS-for-dyn-syst}) applied to the
solution $\Gamma_{k}$ of the flow equation is
\begin{eqnarray}
\mu\frac{d}{d\mu}\Gamma_{k}\left[\varphi\right] & = & \left(\mu\partial_{\mu}+\tilde{\beta}_{i}\left(\tilde{g}^{\left(R\right)}\right)\frac{\partial}{\partial\tilde{g}_{i}^{\left(R\right)}}+\partial_{\log\mu}Z^{\left(R\right)}\frac{\partial}{\partial Z^{\left(R\right)}}\right)\Gamma_{k}\left[\varphi\right]=0  \, , \label{eq:RG-invariance-1}
\end{eqnarray}
where $\partial_{\log\mu}Z^{\left(R\right)}\equiv\partial_{t}Z_{k}\left(\mu\right)$.
Recalling that in the ansatz (\ref{eq:LPAprime_ansatz_Zevery_field})
the field is always accompanied by a factor $Z_{k}^{1/2}$ we rewrite
equation (\ref{eq:RG-invariance-1}) as follows
\begin{eqnarray}
0 & = & \left(\mu\partial_{\mu}+\tilde{\beta}_{i}\left(\tilde{g}^{\left(R\right)}\right)\frac{\partial}{\partial\tilde{g}_{i}^{\left(R\right)}}+\frac{1}{2}\partial_{\log\mu}Z^{\left(R\right)} \left( Z^{\left(R\right)} \right)^{-1}
\varphi\cdot\frac{\delta}{\delta\varphi}\right)\Gamma_{k}\left[\varphi\right]\,.\label{eq:RG-invariance-2}
\end{eqnarray}
In going from (\ref{eq:RG-invariance-1}) to (\ref{eq:RG-invariance-2})
we have been able to trade $\partial/\partial Z^{\left(R\right)}$
with $\delta/\delta\varphi$ by exploiting the fact that the flow
equation for the wave function renormalization has the form 
\begin{eqnarray*}
\partial_{t}Z_{k} & = & f\left(\tilde{g}\right)Z_{k}
\end{eqnarray*}
and the associated solution is:
\begin{eqnarray}
Z_{k}^{\left(sol\right)} & = & Z^{\left(R\right)}\exp\left[\int_{\mu}^{k}f\left(\tilde{g}\left(k^{\prime}\right)\right)\frac{dk^{\prime}}{k^{\prime}}\right]\,.\label{eq:Zsol-explicit-any-k}
\end{eqnarray}
This tells us that for the wave function renormalization the solution
$Z_{k}^{\left(sol\right)}$ is proportional to the boundary condition
$Z^{\left(R\right)}$ implying that the previous rewriting is correct. 

If we think of the EAA as the sum of all the proper vertices,
\begin{eqnarray*}
\Gamma_{k} & = & \sum_{n}\frac{\Gamma_{k}^{\left(n\right)}}{n!}\varphi^{n} \, ,
\end{eqnarray*}
and we functionally derive equation (\ref{eq:RG-invariance-2}) $n$
times with respect to $\varphi$ before setting $\varphi=0$ we obtain:
\begin{eqnarray}
\left(\mu\partial_{\mu}+\beta_{i}^{\left(R\right)}\frac{\partial}{\partial\tilde{g}_{i}^{\left(R\right)}}-\frac{n}{2}\left(- \left(Z^{\left(R\right)}\right)^{-1}\partial_{\log\mu}Z^{\left(R\right)}\right)\right)\Gamma_{k}^{\left(n\right)} & = & 0\,.\label{eq:RG-invariance-CSeq-1}
\end{eqnarray}
Equation (\ref{eq:RG-invariance-CSeq-1}) looks exactly like the Callan-Symanzik
equation of usual quantum field theory once $k\rightarrow0$. 

Let us now discuss in some detail the meaning of equation (\ref{eq:RG-invariance-CSeq-1}) 
in the limit $k\rightarrow0$ in presence of an IR fixed point. In
the fixed point regime the dimensionless couplings $\tilde{g}$ tend
to a constant $\tilde{g}_{*}$ while the equation for the wave function
renormalization takes the simple form
\begin{eqnarray}
\partial_{t}Z_{k} & = & f\left(\tilde{g}_{*}\right)Z_{k}\,.\label{eq:Z-running-FP-regime}
\end{eqnarray}
Let us impose the boundary condition in the fixed point regime, i.e.~$\mu$ is enough small
that the running of $Z_k$ is given approximatively by equation (\ref{eq:Z-running-FP-regime})
(we will come back to an arbitrary $\mu$ in a moment). 
In this limit the solution of the equation
(\ref{eq:Z-running-FP-regime}) is particularly simple and reads:
\begin{eqnarray}
Z_{k}^{\left(sol\right)} & = & Z^{\left(R\right)}\left(\frac{k}{\mu}\right)^{-\eta}\label{eq:Zsol-explicit-FP-regime}
\end{eqnarray}
where $\eta=-f\left(\tilde{g}_{*}\right)$. As we will see in a moment $\eta/2$ can 
be identified with the anomalous dimension of the field at the fixed point. 
In the limit $k\rightarrow0$ all the couplings $\tilde{g}\left(k\right)$ appearing in the EAA tend to a constant
and they no longer depend on $\mu$. 
However, there is still some $\mu-$dependence in the
EAA coming from the wave function renormalization. In this case the
invariance under changes in the normalization point $\mu$ tells us that
\begin{eqnarray}
\left(\mu\partial_{\mu}-\frac{n}{2}\eta\right)\Gamma_{k\rightarrow0}^{\left(n\right)}\left[\varphi\right] & = & 0\,.\label{eq:RG-invariance-CSeq-2-FP}
\end{eqnarray}
This suggests to identify the anomalous dimension at the fixed point
with $\eta=-Z_{k}^{-1}\partial_{t}Z_{k}$ for $k\rightarrow0$. Indeed
using (\ref{eq:RG-invariance-CSeq-2-FP}) and dimensional analysis
for a given proper vertex, i.e.
\begin{eqnarray}
\left(\mu\partial_{\mu}+p\partial_{p} + n\frac{d-2}{2}\right)\Gamma_{k\rightarrow0}^{\left(n\right)}\left[\varphi\right] & = & 0\,, \label{eq:dim_anal_GammaN}
\end{eqnarray}
one can deduce the $\left(p,\mu\right)-$dependence of $\Gamma_{k\rightarrow0}^{\left(n\right)}$ (in (\ref{eq:dim_anal_GammaN}) we consider
functional derivative with respect to the Fourier transform of the field). 
For instance if we consider $\Gamma_{k\rightarrow0}^{\left(2\right)}$ and factor out the overall delta function
entailing momentum conservation we obtain:
\begin{eqnarray*}
\Gamma^{\left(2\right)}\left[\varphi\right] & \sim & p^{2-\eta}\,.
\end{eqnarray*}
This shows that $\eta$ can indeed be identified with the anomalous dimension.

At this point the reader may wonder if the fact of having chosen
boundary conditions for an arbitrarily small $\mu$ played any
role in our arguments. The answer is no, it is by no means necessary
to impose boundary conditions for $\mu$ arbitrarily small, this just
leads to a quicker argument. 
To see this, let us denote $Z^{(R)}$ the boundary condition imposed at some other scale $\mu$.
Again in the fixed point regime the
couplings in the EAA do not depend any longer on the scale $\mu$
or on the boundary condition $\tilde{g}^{\left(R\right)}$. However
now the wave function renormalization has no longer the simple structure
of equation (\ref{eq:Zsol-explicit-FP-regime}). In particular the
solution $Z_{k}^{\left(sol\right)}$ depends also on the boundary
condition $\tilde{g}^{\left(R\right)}$ since the integral in the
exponent of (\ref{eq:Zsol-explicit-any-k}) depends on it. Therefore,
for $k\rightarrow0$ we have a solution of the following form:
\begin{eqnarray}
Z_{k}^{\left(sol\right)} & = & Z^{\left(R\right)}y\left(g^{\left(R\right)}\right)\left(\frac{k}{\mu}\right)^{-\eta} \label{eq:Zsol_expl_k_to_zero}
\end{eqnarray}
for some function $y\left(g^{\left(R\right)}\right)$\footnote{More formally this function can be defined via $y\left(g^{\left(R\right)}\right)\equiv
\lim_{k\rightarrow0} \left(Z^{\left(R\right)} \right)^{-1} \left(\frac{k}{\mu}\right)^{-f\left(\tilde{g}_{*}\right)}Z_{k}^{\left(sol\right)}$.}.
The invariance under floating normalization point transformations tells
us that
\begin{eqnarray*}
\left(\mu\partial_{\mu}+\beta_{i}^{\left(R\right)}\frac{\partial}{\partial\tilde{g}_{i}^{\left(R\right)}}-\frac{n}{2}\left(-
\left( Z^{\left(R\right)} \right)^{-1} \partial_{\log\mu}Z^{\left(R\right)}\right)\right)\Gamma_{k\rightarrow0}^{\left(n\right)} & = & 0 \, ,
\end{eqnarray*}
where the derivatives with respect to the couplings $\tilde{g}^{\left(R\right)}_i$
act only on $Z_{k}^{\left(sol\right)}$ since all the other couplings are approaching their fixed point values. 
Exploiting this we can rewrite the above equation as
\begin{eqnarray*}
\left(\mu\partial_{\mu}+\frac{n}{2}\beta_{i}^{\left(R\right)}\frac{\partial y\left(\tilde{g}^{\left(R\right)}\right)}{\partial\tilde{g}_{i}^{\left(R\right)}}\frac{1}{y\left(\tilde{g}^{\left(R\right)}\right)}-\frac{n}{2}\left(-
\left( Z^{\left(R\right)} \right)^{-1} \partial_{\log\mu}Z^{\left(R\right)}\right)\right)\Gamma_{k\rightarrow0}^{\left(n\right)} & = & 0 \, ,
\end{eqnarray*}
or equivalently as
\begin{eqnarray*}
\left(\mu\partial_{\mu}+\frac{n}{2}\left[\beta_{i}^{\left(R\right)}\frac{\partial y\left(\tilde{g}^{\left(R\right)}\right)}{\partial\tilde{g}_{i}^{\left(R\right)}}\frac{1}{y\left(\tilde{g}^{\left(R\right)}\right)}+f\left(\tilde{g}^{\left(R\right)}\right)\right]\right)\Gamma_{k\rightarrow0}^{\left(n\right)} & = & 0\,.
\end{eqnarray*}
The above equation tells us that we can express the anomalous dimension also as follows:
\begin{eqnarray}
-\eta & = & \beta_{i}^{\left(R\right)}\frac{\partial y\left(\tilde{g}^{\left(R\right)}\right)}{\partial\tilde{g}_{i}^{\left(R\right)}}\frac{1}{y\left(\tilde{g}^{\left(R\right)}\right)}+f\left(\tilde{g}_{i}^{\left(R\right)}\right)\,.\label{eq:Eta-via-arbitrary-mu}
\end{eqnarray}
Although not obvious, the RHS of equation (\ref{eq:Eta-via-arbitrary-mu})
must be independent of $\tilde{g}^{\left(R\right)}$. We checked relation
(\ref{eq:Eta-via-arbitrary-mu}) in few examples.\footnote{
The fact that the RHS of eq.~(\ref{eq:Eta-via-arbitrary-mu}) is just a constant can be understood applying the
$\mu-$invariance operator directly on $Z_{k\rightarrow 0}^{(sol)}$: the explicit $\mu$ dependence comes solely from the factor $\mu^{\eta}$ in 
(\ref{eq:Zsol_expl_k_to_zero}) and, writing down all the terms, one finds eq.~(\ref{eq:Eta-via-arbitrary-mu}).} 
It is clear now
why choosing the boundary condition with $\mu$ arbitrarily small
is convenient: if we let $\mu\rightarrow0$ the first term on the RHS of (\ref{eq:Eta-via-arbitrary-mu})
vanishes and we are left with $-\eta=f\left(\tilde{g}_{*}\right)=Z_{k}^{-1}\partial_{t}Z_{k}|_{k=0}$.
Thus our arguments suggest to identify the anomalous dimension with $-Z_{k}^{-1}\partial_{t}Z_{k}$ in the limit $k\rightarrow0$.

With respect to the standard Callan-Symanzik equation we derived equation
(\ref{eq:RG-invariance-CSeq-1}) using dimensionless couplings. However,
nothing prevents us from repeating the same reasoning for the dimensionful couplings. 
It is just more convenient to work with dimensionless quantities
since those are the ones of interest in the fixed
point regime of the EAA. Moreover, if one repeats our reasoning in the case of
dimensionful couplings, one notes that equation (\ref{eq:dim_anal_GammaN})
involves new terms of the type $d_g g^{(R)}\partial / \partial g^{(R)}$ where $d_g$
is the mass dimension of the coupling.
After putting everything together one notices that it is just the dimensionless beta function which enters 
in the scaling equation once $\mu$ has been eliminated.
Equations describing $\mu-$invariance in a somewhat different manner are also known in the Wilson-Polchinski framework, see in particular
\cite{Sonoda:2006ai} for a discussion including a choice of parametrization scheme related to
the MS scheme in dimensional regularization.

A further motivation for the identification of $\eta=-Z_{k}^{-1}\partial_{t}Z_{k}|_{k=0}$ as the anomalous dimension
has been provided in \cite{Blaizot:2005wd}. Thus let us briefly consider how the
arguments in appendix A of \cite{Blaizot:2005wd} apply to our framework since
we will employ similar arguments in the next section to provide a
further reason in favour of our definitions. Dimensional analysis tells us
that:
\begin{eqnarray*}
\frac{\Gamma_{k}^{\left(2\right)}\left(p,k,\mu\right)}{\Gamma_{k}^{\left(2\right)}\left(p^{\prime},k,\mu\right)} & = & \hat{f}\left(\frac{p^{\prime}}{p},\frac{p}{k},\frac{p^{\prime}}{\mu}\right)\,.
\end{eqnarray*}
The arguments of $\hat{f}$ are three possible independent ratios,
all the other ratios can be obtained from them. 
Now we take $k$ to be sufficiently small (fixed point regime) and observe that in the EAA the $\mu$ dependence is contained
only in the wave function renormalization factors. 
Thus, in the above ratio, these wave function renormalizations cancel against each other
and we have
\begin{eqnarray*}
\frac{\Gamma_{k}^{\left(2\right)}\left(p,k,\mu\right)}{\Gamma_{k}^{\left(2\right)}\left(p^{\prime},k,\mu\right)} & = & \hat{f}\left(\frac{p^{\prime}}{p},\frac{p}{k} \right)\,.
\end{eqnarray*}
Setting $p^{\prime}=0$ we obtain:
\begin{eqnarray*}
\Gamma_{k}^{\left(2\right)}\left(p,k,\mu\right) & = & \Gamma_{k}^{\left(2\right)}\left(0,k,\mu\right)\hat{f}\left(0,\frac{p}{k} \right)\,.
\end{eqnarray*}
Let us denote $f\left(\frac{p}{k}\right)\equiv\hat{f}\left(0,\frac{p}{k} \right)$
and observe that in the fixed point regime
\begin{eqnarray*}
\Gamma_{k}^{\left(2\right)}\left(0,k,\mu\right) & \sim & k^{2}\left(\frac{k}{\mu}\right)^{-\eta}\,.
\end{eqnarray*}
Therefore we finally have:
\begin{eqnarray*}
\Gamma_{k}^{\left(2\right)}\left(p,k,\mu\right) & \sim & k^{2}\left(\frac{k}{\mu}\right)^{-\eta}f\left(\frac{p}{k}\right)\,.
\end{eqnarray*}
In order for this expression to be well defined in the critical regime,
i.e.~$k=0$, we require that $f\sim\left(p/k\right)^{2-\eta}$ and
thus
\begin{eqnarray*}
\Gamma_{0}^{\left(2\right)}\left(p,0,\mu\right) & \sim & p^{2}\left(\frac{p}{\mu}\right)^{-\eta}\,.
\end{eqnarray*}
This completes the argument to interpret $\eta$ as the anomalous dimension.
These considerations also provide a justification for the argument
telling us that $\Gamma_{k}^{\left(2\right)}\left(p=k\right)\sim k^{2-\eta}$,
which is sometime used in the FRG literature.

\subsection{Fixed point and anomalous dimension of composite operators \label{sub:Fixed-point-and-anomalous-dim}}

In this section we extend our reasoning to the modified EAA $\Gamma_{k}\left[\varphi,\varepsilon\right]$
introduced in section \ref{sec:RG-flow-of-composite-operators}. We
are interested in the scaling of composite operators at a fixed point.
Let us consider a composite operator parametrized via a sum of various
operators. To gain some insights we shall consider perturbation theory
as a first hint. We parametrize a renormalized composite operator of mass dimension
$d_{O}$ with the sum of all possible composite operators of mass
dimension smaller or equal to $d_{O}$. For instance in a six dimensional
$\varphi^{3}$ theory one can consider \cite{Collins:1984xc}:
\begin{eqnarray}
\left[\varphi^{2}\right] & = & Z_{a}\frac{\varphi^{2}}{2}+\left(Z_{b}m^{2}\right) \varphi +Z_{c}\Delta\varphi\,.\label{eq:composite-operator-phi2-collins}
\end{eqnarray}
To obtain a complete information one should define simultaneously all
the composite operators which form the basis $\left\{ O_{i}\right\} $:
\begin{eqnarray*}
\left[O_{i}\right] & = & Z_{ij}O_{j}\,.
\end{eqnarray*}
For instance in the example of the operator $\left[\varphi^{2}\right]$
one should consider \cite{Collins:1984xc}\footnote{The known scale dependence of the one loop matrix elements can be
easily computed via eq.~(\ref{eq:flow_eq_composite_operator_epsilon}).}:
\begin{eqnarray*}
\left(\begin{array}{c}
\left[\frac{1}{2}\varphi^{2}\right]\\
\left[\varphi\right]\\
\left[\Delta\varphi\right]
\end{array}\right) & = & \left(\begin{array}{ccc}
Z_{a} & Z_{b}m^{2} & Z_{c}\\
0 & 1 & 0\\
0 & 0 & 1
\end{array}\right)\left(\begin{array}{c}
\frac{1}{2}\varphi^{2}\\
\varphi\\
\Delta\varphi
\end{array}\right)\,.
\end{eqnarray*}
Let us note that the entries of $Z_{ij}$ are in general dimensionful.
In dimensional regularization dimensionful factors of $Z_{ij}$ are
due to the presence of dimensionful couplings like the mass.
In our scheme, however, it must be generically expected that some dimensionful
factors may depend on the scale $k$ itself.

Parametrizations of composite operators similar to the ones in equations
(\ref{eq:composite_operator_example_ItkZub}) and (\ref{eq:composite-operator-phi2-collins})
can be straightforwardly adopted in our flow equation (\ref{eq:flow_eq_composite_operator_epsilon}).
The difference with the standard framework is due to the fact that, in our scheme, closed families of operators under renormalization are
generically infinite dimensional and not just a finite set like it
happens when using dimensional regularization. To make progress let
us consider the flow equation for composite operators where we insert
(in)finitely many operators adding $\varepsilon_{i}Z_{ij}O_{j}$ to
the EAA. After denoting $G_{k}$ the regularized inverse propagator,
\begin{equation*}
G_k \equiv \left(\Gamma_{k}^{\left(2\right)}+R_{k}\right)^{-1} \, ,
\end{equation*}
we rewrite equation (\ref{eq:flow_eq_composite_operator_epsilon}) as
\begin{eqnarray}
\partial_{t}\left(\varepsilon_{i}Z_{ij}O_{j}\right) & = & -\frac{1}{2}G_{k}\cdot\left(\varepsilon_{i}Z_{ij}O_{j}^{\left(2\right)}\right)\cdot G_{k}\cdot\partial_{t}R_{k}\,.\label{eq:flow_composite_operator_epsilon_Zij}
\end{eqnarray}

Now we want to extend to $\Gamma_{k}\left[\varphi,\varepsilon\right]$
equations (\ref{eq:RG-invariance-2}) and (\ref{eq:RG-invariance-CSeq-1}).
This will allow us to identify which are the scaling dimensions
of the composite operators. The reasoning of section \ref{sub:RG-invariance-of-fRG}
applies straightforwardly to the $\varepsilon$ dependent EAA. Let
$\left\{ O_{i}\right\} $ be the set of operators which parametrize
the composite operators. In analogy with the example (\ref{eq:composite_operator_example_ItkZub})
a renormalized composite operator is defined as
\begin{eqnarray*}
\left[O_{i}\right]\left(\varphi\right) & = & Z_{ij}O_{j}\left(Z_{k}^{1/2}\varphi\right)\,.
\end{eqnarray*}
Since we shall be interested in just one insertion of a composite
operator we will limit ourselves to consider sources $\varepsilon$
such that $\varepsilon^{2}=0$. 
In this case we can write
\begin{eqnarray*}
\Gamma_{k}\left[\varphi,\varepsilon\right] & = & \Gamma_{k}\left[\varphi\right]+\varepsilon\cdot Z\cdot O\,.
\end{eqnarray*}
Let $Z_{ij}^{\left(R\right)}$ be the boundary condition associated
to the flow of the mixing matrix $Z_{ij}$. Applying the reasoning
of section \ref{sub:RG-invariance-of-fRG} and treating $Z_{ij}^{\left(R\right)}$
as all the other boundary conditions we obtain:
\begin{eqnarray*}
\mu\frac{d}{d\mu}\Gamma_{k}\left[\varphi\right] & = & \left(\mu\partial_{\mu}+\beta_{i}^{\left(R\right)}\frac{\partial}{\partial\tilde{g}_{i}^{\left(R\right)}}+\partial_{\log\mu}Z^{\left(R\right)}\frac{\partial}{\partial Z^{\left(R\right)}}+\partial_{\log\mu}Z_{ij}^{\left(R\right)}\cdot\frac{\partial}{\partial Z_{ij}^{\left(R\right)}}\right)\Gamma_{k}\left[\varphi,\varepsilon\right]=0 \, ,
\end{eqnarray*}
where $\partial_{\log\mu}Z_{ij}^{\left(R\right)}=\partial_{t}Z_{ij}\left(k=\mu\right)$.
In section \ref{sub:RG-invariance-of-fRG} we have been able to trade
the partial derivative with respect to $Z^{\left(R\right)}$ for
a functional derivative with respect to the field. Here we would like to do something
similar and trade the derivative with respect to $Z_{ij}^{\left(R\right)}$
for a functional derivative with respect to the source $\varepsilon_i$. In section
\ref{sub:RG-invariance-of-fRG} this step was straightforward since
the solution of the wave function renormalization $Z_{k}^{\left(sol\right)}$
is proportional to $Z^{\left(R\right)}$. The situation for the solution
of the mixing matrix $Z_{ij}^{\left(sol\right)}$ is slightly more complicated.
Taking a functional derivative with respect to $\varepsilon_{i}$
in equation (\ref{eq:flow_composite_operator_epsilon_Zij}) one obtains
that the flow of the mixing matrix has the following form:
\begin{eqnarray}
Z_{ij}^{-1}\partial_{t}Z_{jk} & = & f_{ik}\left(\tilde{g},k\right)\,. \label{eq:flow_eq_Zij}
\end{eqnarray}
The solution of this type of equation is related to a $k-$ordered exponential
matrix (Dyson's series). The crucial point, however, is the following: a generic boundary
condition $Z_{ij}^{\left(R\right)}$ appears in the solution via $Z_{ij}^{\left(sol\right)}=Z_{im}^{\left(R\right)}M_{mj}$
for some matrix $M_{mj}$. This fact allows us to trade the derivative
$\partial/\partial Z_{ij}^{\left(R\right)}$ with a suitable functional
derivative with respect to the source $\varepsilon$. In particular we can rewrite:
\begin{eqnarray}
\mu\frac{d}{d\mu}\Gamma_{k}\left[\varphi , \varepsilon \right] & = & 
\Bigr[
\mu\partial_{\mu}+\beta_{i}^{\left(R\right)}\frac{\partial}{\partial\tilde{g}_{i}^{\left(R\right)}}
+\frac{1}{2} \frac{\partial_{\log\mu}Z^{\left(R\right)}}{Z^{\left(R\right)}}  \varphi\cdot\frac{\delta}{\delta\varphi} 
+\varepsilon_{i}\cdot\gamma_{Z,ij}\cdot\frac{\delta}{\delta\varepsilon_{j}} \Bigr] 
\Gamma_{k}\left[\varphi,\varepsilon\right]=0 \, , \qquad  \label{eq:mu_inv_Gamma_phi_eps}
\end{eqnarray}
where 
\begin{equation*}
\gamma_{Z,ik}\equiv\partial_{\log\mu}Z_{ij}^{\left(R\right)} \left(Z^{\left(R\right)}\right)^{-1}_{jk} \, . 
\end{equation*}
Let us impose the boundary condition in the fixed point regime, i.e.~at $\mu$ small enough that the running of $Z_{ij}$ is 
given by equation (\ref{eq:flow_eq_Zij}) with $\tilde{g}=\tilde{g}_*$.
Then, using the result of section \ref{sub:RG-invariance-of-fRG} for $\eta$, 
in the fixed point regime we can rewrite (\ref{eq:mu_inv_Gamma_phi_eps}) as:
\begin{eqnarray}
\mu\frac{d}{d\mu}\Gamma_{k}\left[\varphi,\varepsilon\right] & = & \left(\mu\partial_{\mu}-\frac{1}{2}\eta\varphi\cdot\frac{\delta}{\delta\varphi}+\varepsilon_{i}\cdot\gamma_{Z,ij}\cdot\frac{\delta}{\delta\varepsilon_{j}}\right)\Gamma_{k\rightarrow0}\left[\varphi,\varepsilon\right]=0 \, .  \label{eq:mu_inv_Gamma_phi_eps_FP}
\end{eqnarray}

In order to discuss the scaling associated to composite operators we need to take into account 
both equation (\ref{eq:mu_inv_Gamma_phi_eps_FP}) and dimensional analysis.
This is fully analogous to what we did in section \ref{sub:RG-invariance-of-fRG} for the
wave function renormalization.
However, there is now a crucial difference, namely the fact that the boundary condition $Z_{ij}^{\left(R\right)}$
are now dimensionful parameters which must be taken into account in the dimensional analysis. 
Let $d_i$ be the mass dimension of the operator $O_i$, then the matrix element $Z_{ij}^{\left(R\right)}$ 
has mass dimension $\left(d_{i}-d_{j}\right)$ and,
focusing our attention only on one insertion of a composite
operator, i.e.~taking a single functional derivative with respect to the source $\varepsilon_{i}$
and setting $\varepsilon_{i}$ to zero, we obtain:
\begin{eqnarray*}
\left(\mu\partial_{\mu}-x\partial_{x}+\left(d_{i}-d_{j}\right)Z_{ij}^{\left(R\right)}\frac{\partial}{\partial Z_{ij}^{\left(R\right)}}-d_{i}\delta_{ij}\right)\frac{\delta\Gamma_{k\rightarrow0}}{\delta\varepsilon_{i}}  & = & 0\, ,
\end{eqnarray*}
where the last term takes into account the mass dimension of the operator $O_i$.
Introducing the matrix
\begin{equation*}
D_{ij}=d_i \delta_{ij} \, ,
\end{equation*}
we can rewrite the third term in the brackets via $\left(d_{i}-d_{j}\right)Z_{ij}^{\left(R\right)}=D_{ia}Z_{aj}^{\left(R\right)}-Z_{ia}^{\left(R\right)}D_{aj}$
and trade the derivative of $Z_{ij}^{\left(R\right)}$ for $\left(Z_{ij}^{\left(R\right)}\right)^{-1}$.
In this manner we obtain (in matrix notation):
\begin{eqnarray*}
\left(\mu\partial_{\mu}-x\partial_{x}+\left(DZ^{\left(R\right)}-Z^{\left(R\right)}D\right)\left(Z^{\left(R\right)}\right)^{-1}-D\right)\frac{\delta\Gamma_{k\rightarrow0}}{\delta\varepsilon_{i}}  & = & 0\\
\left(\mu\partial_{\mu}-x\partial_{x}-Z^{\left(R\right)}D\left(Z^{\left(R\right)}\right)^{-1}\right)\frac{\delta\Gamma_{k\rightarrow0}}{\delta\varepsilon_{i}}  & = & 0\,.
\end{eqnarray*}
Now we eliminate $\mu$ from the above equation using (\ref{eq:mu_inv_Gamma_phi_eps_FP}) and we obtain:
\begin{eqnarray}
\left(x\partial_{x}+Z^{\left(R\right)}D\left(Z^{\left(R\right)}\right)^{-1}+\gamma_{Z}\right)\frac{\delta\Gamma_{k\rightarrow0}}{\delta\varepsilon_{i}}
& = & 0\,. \label{eq:CS_plus_dimAnalysis}
\end{eqnarray}
The eigenvalues of the matrix $Z^{\left(R\right)}D\left(Z^{\left(R\right)}\right)^{-1}+\gamma_{Z}$
yield the full, i.e.~classical plus anomalous, dimensions of the
scaling operators. To see this let us assume that $Z^{\left(R\right)}D\left(Z^{\left(R\right)}\right)^{-1}+\gamma_{Z}$
is diagonalizable, i.e.~$Z^{\left(R\right)}D\left(Z^{\left(R\right)}\right)^{-1}+\gamma_{Z}=A E A^{-1}$
where $E_{ab}=e_{a}\delta_{ab}$ is the eigenvalue matrix. Then we
can manipulate equation (\ref{eq:CS_plus_dimAnalysis}) as follows:
\begin{eqnarray*}
\left(x\partial_{x}\delta_{ij}+A_{ia}E_{ab}A_{bj}^{-1}\right)\frac{\delta\Gamma_{k\rightarrow0}}{\delta\varepsilon_{j}} & = & 0\\
\left(x\partial_{x}+e_{b}\right)A_{bj}^{-1}\frac{\delta\Gamma_{k\rightarrow0}}{\delta\varepsilon_{j}}  & = & 0\,.
\end{eqnarray*}
The last equation tells us that $A_{bj}^{-1}\left[O_{j}\right]$
is a scaling operator and
we can identify $e_{b}$ with its full dimension, which we denote $\Delta_{b}$. 
In analogy with the case of the wave function renormalization, 
this shows that the full dimensions of the scaling operators are given by the eigenvalues of the matrix
$ZD Z^{-1}+\gamma_{Z}$ in the limit $k\rightarrow 0$. 

It turns out convenient
to consider the dimensionless analogue of $Z_{ij}$, which we denote $N_{ij}$.
Introducing the matrix 
\begin{eqnarray}
K_{ij} & = & k^{d_{i}}\delta_{ij}\,,\,\, d_{i}\equiv\mbox{mass dimension of }O_{i}\left(\varphi\right)  \label{eq:def_matrix_K}
\end{eqnarray}
the mixing matrix $Z_{ij}$ can be rewritten via a dimensionless
matrix $N_{ij}$ defined as follows 
\begin{eqnarray*}
N_{il} & \equiv & K_{ij}^{-1}Z_{jk}K_{kl}\,.
\end{eqnarray*}
In particular we want to show that the spectrum of $Z DZ^{-1}+\gamma_{Z}$
is the same as $D+\gamma_{N}$, where $\gamma_N=\partial_t N N^{-1}$.
First we observe that:
\begin{eqnarray*}
ZDZ^{-1}+\gamma_{Z} & = & ZDZ^{-1}+\partial_{t}Z Z^{-1}\\
 & = & Z\left(D+Z^{-1}\partial_{t}Z\right) Z^{-1}\,.
\end{eqnarray*}
Thus the matrix $ZDZ^{-1}+\gamma_{Z}$
has the same spectrum as $D+Z^{-1}\partial_{t}Z$.
Now we shall check that indeed these matrices yield the same scaling dimensions,
i.e.~same spectrum, as $D+\partial_{t}N N^{-1}$.
In matrix notation we have
\begin{eqnarray*}
Z^{-1}\partial_{t}Z & = & \left(KN^{-1}K^{-1}\right)\partial_{t}\left(KNK^{-1}\right)\\
 & = & Z^{-1}DZ+KN^{-1}\partial_{t}NK^{-1}-D\\
 & = & Z^{-1}K\left(D+{\gamma}_{N}\right)K^{-1}Z-D \, ,
\end{eqnarray*}
from which our claim follows. Similarly, the matrix $N^{-1}DN+N^{-1}\partial_{t}N $
has the same spectrum as $D+\gamma_N$.

All these manipulations are quite formal and we would like to provide
a more intuitive argument for our definition. Let us consider the
composite operator $\left[\varphi^{2}\right]$ and parametrize it
via the following simple ansatz $\left[\varphi^{2}\right]=Z_{\varphi^{2}}Z_{k}\varphi^{2}$.
Now we shall follow the arguments used at the end of section \ref{sub:RG-invariance-of-fRG}.
Let us consider the dimensionless quantity $\Gamma_{k}^{\left(2,1\right)}$,
where in the superscript we indicated the number of functional derivatives
with respect to $\varphi$ and $\varepsilon$ respectively. $\Gamma_{k}^{\left(2,1\right)}$
satisfies:
\begin{eqnarray*}
\Gamma_{k}^{\left(2,1\right)} & = & \frac{G_{k}^{\left(2,1\right)}\left(p_{1},p_{2}\right)}{G_{k}\left(p_{1}\right)G_{k}\left(p_{2}\right)} 	\, ,
\end{eqnarray*}
where
\begin{eqnarray*}
G_{k}^{\left(2,1\right)}\left(p_{1},p_{2}\right) & = & \int dx_{1}dx_{2} dy \,e^{ip_{1}x_{1}+ip_{2}x_{2}}
\langle\left[\varphi^{2}\left(y\right)\right]\varphi\left(x_{1}\right)\varphi\left(x_{2}\right)\rangle\,.
\end{eqnarray*}
We consider $p=p_{1}=-p_{2}$ and observe that in the fixed point regime:
\begin{eqnarray*}
\frac{\Gamma_{k}^{\left(2,1\right)}\left(p,k,\mu\right)}{\Gamma_{k}^{\left(2,1\right)}\left(p^{\prime},k,\mu\right)} & = & \hat{f}\left(\frac{p^{\prime}}{p},\frac{p}{k}\right)\, ,
\end{eqnarray*}
where the RHS do not depend on $\mu$ since the various wave functions renormalizations 
cancel against each other in the ratio.
Once again we have:
\begin{eqnarray*}
\Gamma_{k}^{\left(2,1\right)}\left(p,k,\mu\right) & = & \Gamma_{k}^{\left(2,1\right)}\left(p^{\prime},k,\mu\right)\hat{f}\left(\frac{p^{\prime}}{p},\frac{p}{k}
\right)
\end{eqnarray*}
and setting $p^{\prime}=0$ we obtain:
\begin{eqnarray*}
\Gamma_{k}^{\left(2,1\right)}\left(p,k,\mu\right) & = & \Gamma_{k}^{\left(2,1\right)}\left(0,k,\mu\right)\hat{f}\left(0,\frac{p}{k} \right)\,.
\end{eqnarray*}
Let us denote $f\left(\frac{p}{k}\right)\equiv\hat{f}\left(0,\frac{p}{k}\right)$
and $\gamma_{\varphi^{2}}=Z_{\varphi^{2}}^{-1}\partial_{t}Z_{\varphi^{2}}$.
We observe that in the fixed point regime
\begin{eqnarray*}
\Gamma_{k}^{\left(2,1\right)}\left(0,k,\mu\right) & \sim & \left(\frac{k}{\mu}\right)^{\gamma_{\varphi^{2}}-\eta}\,.
\end{eqnarray*}
Repeating the reasoning of section \ref{sub:RG-invariance-of-fRG}
we obtain:
\begin{eqnarray*}
\Gamma_{0}^{\left(2,1\right)}\left(p,0,\mu\right) & \sim & \left(\frac{p}{\mu}\right)^{\gamma_{\varphi^{2}}-\eta}\,.
\end{eqnarray*}
Thus we identify $\gamma_{\varphi^{2}}=Z_{\varphi^{2}}^{-1}\partial_{t}Z_{\varphi^{2}}$
with the anomalous dimension of $\left[\varphi^{2}\right]$ as expected.

\subsection{Scaling dimensions from the flow equation \label{sub:Scaling-dimension-from-flow-eq}}

In the previous section we deduced the quantities which define
the full dimension of scaling operators. Here we shall show how these
quantities are most easily computed in the FRG framework. Let us recall
that at the fixed point it is convenient to work via dimensionless
objects which are defined using the cutoff and suitable rescalings.
In particular we define:
\begin{eqnarray*}
\varphi\left(x\right) & = & \tilde{\varphi}\left(\tilde{x}\right)k^{\frac{d-2}{2}}\\
\varepsilon_{i}\left(x\right) & = & \tilde{\varepsilon}_{i}\left(\tilde{x}\right)k^{d-d_{i}}\\
x & = & \tilde{x}k^{-1}\,.
\end{eqnarray*}
Being $\varphi\left(x\right)$, $\varepsilon\left(x\right)$
and $x$ independent of the scale $k$ by definition, we obtain:
\begin{eqnarray*}
\partial_{t}\tilde{\varphi}\left(\tilde{x}\right) & = & -\left(\frac{d-2}{2}\right)\tilde{\varphi}\left(\tilde{x}\right)\\
\partial_{t}\tilde{\varepsilon}_{i}\left(\tilde{x}\right) & = & -\left(d-d_{i}\right)\tilde{\varepsilon}_{i}\left(\tilde{x}\right)\\
\partial_{t}\tilde{x} & = & \tilde{x}\,.
\end{eqnarray*}
An operator $O=\partial_{x}^{m}\varphi^{n}$ will satisfy
\begin{eqnarray*}
\partial_{t}\left(\partial_{x}^{m}\varphi^{n}\right) & = & 0 \, ,
\end{eqnarray*}
implying
\begin{eqnarray*}
\partial_{t}\left(k^{m}k^{\frac{d-2}{2}n}\partial_{\tilde{x}}^{m}\tilde{\varphi}^{n}\right) & = & 0\\
\partial_{t}\left(\partial_{\tilde{x}}^{m}\tilde{\varphi}^{n}\right) & = & -d_{O}\left(\partial_{\tilde{x}}^{m}\tilde{\varphi}^{n}\right)\,,
\end{eqnarray*}
where $d_{O}=m+\frac{d-2}{2}n$. 
For a LPA approximation of $\tilde{O}_{l}$ we can rewrite this last
term also as\footnote{In the LPA$^{\prime}$ this term gets a further contribution coming from the
anomalous dimension. If derivatives are present it might be convenient
to express this term via $\frac{\delta\tilde{O}_{l}}{\delta\tilde{\phi}} \left(-\frac{d-2+\eta}{2}\tilde{\phi}\right) 
-\tilde{p}^{\mu}\frac{\partial\tilde{O}_{l}}{\partial\tilde{p}^{\mu}}$.} 
\begin{eqnarray*}
\partial_{t}\tilde{O}_{l} & = & \frac{\delta\tilde{O}_{l}}{\delta\tilde{\varphi}}\left(-\frac{d-2}{2}\tilde{\varphi}\right)\,.
\end{eqnarray*}
In order to express the flow equation
(\ref{eq:flow_composite_operator_epsilon_Zij}) in terms of dimensionless
variables it is convenient to perform the following manipulation:
\begin{eqnarray*}
\int d^{d}x\varepsilon_{i}Z_{ij}O_{i} & = & \int d^{d}\tilde{x}\tilde{\varepsilon}_{i}K_{ij}^{-1}Z_{jk}K_{kl}\tilde{O}_{l}\\
 & = & \int d^{d}\tilde{x}\tilde{\varepsilon}_{i}N_{il}\tilde{O}_{l}\,.
\end{eqnarray*}
The LHS of the flow equation reads
\begin{equation}
\partial_{t}\int d^{d}x\varepsilon_{i}Z_{ij}O_{i}  =  \int d^{d}\tilde{x}\left[d_{i}\tilde{\varepsilon}_{i}N_{il}\tilde{O}_{l}+\tilde{\varepsilon}_{i}\left(\partial_{t}N_{il}N_{lm}^{-1}\right)N_{mn}\tilde{O}_{n}
-\frac{d-2}{2}\tilde{\varepsilon}_{i}N_{il}\frac{\delta\tilde{O}_{l}}{\delta\tilde{\varphi}} \tilde{\varphi}\right] .
\,\,\,\, \label{eq:LHS_flow_dimLess}
\end{equation}
In the above expression the first term comes from the logarithmic
$k$ derivative of $d^{d}\tilde{x}$ and $\tilde{\varepsilon}_{i}$
and the third term from the derivative acting on $\tilde{O}_{l}$.
In the second term of (\ref{eq:LHS_flow_dimLess}) we inserted an
identity $N^{-1}\cdot N$ in order to make explicit the presence of
$\gamma_{N}=\partial_{t}N\cdot N^{-1}$, which enters in the definition
of scaling dimension as shown in the previous section. At this point
it proves convenient to introduce the new basis of operators $\tilde{B}_{i}=N_{il}\tilde{O}_{l}$
and bring the last term in (\ref{eq:LHS_flow_dimLess}) to the RHS of (\ref{eq:flow_composite_operator_epsilon_Zij}). 
After these manipulations we are left with an equation of the form
\begin{eqnarray}
\int d^{d}\tilde{x}\left[\tilde{\varepsilon}_{i}\left(d_{i}\delta_{ij}+\partial_{t}N_{il}N_{lj}^{-1}\right)\tilde{B}_{j}\right] & = & \int d^{d}\tilde{x}\tilde{\varepsilon}_{i}\frac{\delta\tilde{B}_{i}}{\delta\tilde{\varphi}}\left(\frac{d-2}{2}\tilde{\varphi}\right)+\mbox{Tr} \Bigr[\cdots \Bigr]
\,,\label{eq:flow_eq_dimLess}
\end{eqnarray}
where the last term indicates the RHS of equation (\ref{eq:flow_composite_operator_epsilon_Zij}).
Clearly the RHS of equation (\ref{eq:flow_composite_operator_epsilon_Zij})
is proportional to
\begin{eqnarray*}
N_{il}\frac{\delta^{2}\tilde{O}_{l}}{\delta\tilde{\varphi}^{2}} & = & \frac{\delta^{2}\tilde{B}_{i}}{\delta\tilde{\varphi}^{2}}\,.
\end{eqnarray*}
We observe that the quantity in round brackets on the LHS of (\ref{eq:flow_eq_dimLess})
is precisely the matrix $D+\partial_t N \cdot N^{-1}$, whose eigenvalues are the full dimensions
of the scaling operators. Let us assume that the matrix $d_{i}\delta_{ij}+\partial_{t}N_{il}N_{lj}^{-1}$
is diagonalizable and let us denote $\lambda_{i}$, $\Lambda_{ij}$
and $A_{ij}$ the eigenvalues, the eigenvalue matrix and the eigenvector
matrix respectively. After taking a functional derivative with respect
to the source $\tilde{\varepsilon}_{i}$ we can rewrite equation (\ref{eq:flow_eq_dimLess})
as follows:
\begin{eqnarray}
A_{ia}\Lambda_{ab}A_{bj}^{-1}\tilde{B}_{j} & = & \frac{\delta\tilde{B}_{i}}{\delta\tilde{\varphi}}\left(\frac{d-2}{2}\tilde{\varphi}\right)-\frac{1}{2}G_{k}\cdot\frac{\delta^{2}\tilde{B}_{i}}{\delta\tilde{\varphi}^{2}}\cdot G_{k}\cdot\partial_{t}R_{k}\,,\label{eq:flow_eq_dimLess_RHS_explicit}
\end{eqnarray}
where the last term in the RHS is meant solely to represent schematically
the structure of the ``trace term'' in (\ref{eq:flow_eq_dimLess}). At this point it is convenient to
multiply equation (\ref{eq:flow_eq_dimLess_RHS_explicit}) by $A_{mi}^{-1}$ and introduce a new set of
operators $\tilde{D}_{m}\equiv A_{mi}^{-1}\tilde{B}_{i}$. Writing
explicitly the RHS of (\ref{eq:flow_eq_dimLess_RHS_explicit}) in
the LPA approximation we have:
\begin{eqnarray}
\lambda_{i}D_{i}\left(\tilde{\varphi}\right)& = & D_{i}^{\prime}\left(\tilde{\varphi}\right)\left(\frac{d-2}{2}\tilde{\varphi}\right)  -c_{d}\frac{1}{1+\tilde{U}_{k}^{\prime\prime}\left(\tilde{\varphi}\right)}D_{i}^{\prime\prime}\left(\tilde{\varphi}\right)\frac{1}{1+\tilde{U}_{k}^{\prime\prime}\left(\tilde{\varphi}\right)}\, , \label{eq:lambda_composite_operator_integrated}
\end{eqnarray}
where $c_{d}^{-1}=\left(4\pi\right)^{d/2}\Gamma\left(d/2+1\right)$ (more details are given in section \ref{sec:Scaling-solutions-and-composite-operators}).
Remarkably equation (\ref{eq:lambda_composite_operator_integrated})
is expressed directly in terms of the full dimension $\lambda_{i}$
of the scaling operators and thus its solutions yield directly the
scaling dimensions of the operator content of the fixed point theory.
In section \ref{sec:Scaling-solutions-and-composite-operators} we
will see that, considering composite operators of the form $O\left(\varphi\right)$
within the LPA$^{\prime}$, the eigenvalues $\lambda_{i}$ are directly connected
to the critical exponents $\theta_{i}$ via $\lambda_{i}=d-\theta_{i}$.
Note also that one can also arrive at equation (\ref{eq:lambda_composite_operator_integrated}) by
taking a functional derivative with respecto to $\tilde{\varepsilon}_i$ of (\ref{eq:LHS_flow_dimLess})
and applying the matrix $N^{-1}$ to it, in this case one diagonalizes the matrix $N^{-1}DN+N^{-1} \partial_t N$.

Let us conclude this section by commenting some possible contacts with other works in the literature.
In particular it would be nice to set up our discussion in a geometric language along the lines
considered in \cite{Lassig:1989tc,Sonoda:1993dh,Dolan:1994pq} (see also \cite{Polonyi:2000fv} for a slightly different approach).
In these works, roughly speaking, composite operators are thought of as living in the tangent space associated to the theory space.
The quantity $\gamma_a^b \equiv \partial_{g_a} \beta^b$ naturally appears and one can derive
a Callan-Symanzik type of equation by considering the RG as a one-parameter group of diffeomorphisms \cite{Lassig:1989tc}. 
The matrix $\gamma_a^b$ can be interpreted as the anomalous dimension mixing matrix.
In our approach this can be understood via the following argument. Let us limit ourselves to parametrize composite operators
via operators which are not total derivatives and take the sources $\varepsilon_i$ to be constants. Moreover
let us parametrize the EAA via $\Gamma_k = \sum_i g_i O_i$.
From equation (\ref{eq:flow_composite_operator_epsilon_Zij}) we obtain:
\begin{eqnarray}
Z^{(-1)}_{ia}\left(\partial_{t} Z_{aj}\right)O_{j} & = & -\frac{1}{2} \frac{1}{\Gamma_k^{(2)}+R_k} \cdot\left(O_{i}^{\left(2\right)}\right)\cdot \frac{1}{\Gamma_k^{(2)}+R_k} \cdot\partial_{t}R_{k}\,.  \label{eq:ZinvDZ_geom}
\end{eqnarray}
Now let us write the flow equation (\ref{eq:flow_eq_EAA}) for $\Gamma_k = \sum_i g_i O_i$:
\begin{eqnarray*}
\sum_j \beta^j O_j & = & \frac{1}{2} \frac{1}{\sum_j g_j O_j^{(2)} +R_{k}} \cdot \partial_{t}R_{k} \, .
\end{eqnarray*}
Taking a derivative with respect to $g_i$ we obtain:
\begin{eqnarray}
\sum_j \partial_{g_i} \beta^j O_j & = & -\frac{1}{2} \frac{1}{\sum_j g_j O_j^{(2)} +R_{k}} \cdot O_i^{(2)} \cdot 
\frac{1}{2} \frac{1}{\sum_j g_j O_j^{(2)} +R_{k}} \cdot \partial_{t}R_{k} \, , \label{eq:Dg_flow_geom}
\end{eqnarray}
Comparing (\ref{eq:ZinvDZ_geom}) to (\ref{eq:Dg_flow_geom}) one concludes
that $\gamma_a^b = \partial_{g_{a}} \beta^b$ equals $Z^{(-1)}_{ac} \partial_{t} Z_{cb}$. 
However, let us note once again that, for the above argument to go through, we had to 
neglect some total derivative operators whose contribution might be important. 
Finally more work is needed to spell out the possible geometrical interpretations of 
our arguments, we hope to come back to these issues in the future.

\section{Scaling solutions and composite operators \label{sec:Scaling-solutions-and-composite-operators}}

In this section we consider approximate solutions of the fixed point equation
and how one can use them to estimate the anomalous dimensions of various operators
at the fixed point. 
We parametrize composite operators as functions of the field but not of its derivatives.
As we shall see, this choice can be put in one-to-one correspondence with 
results known within the LPA$^{\prime}$ approximation 
and we shall comment them in view of the discussion of the previous section. 

Scaling solutions are (approximate) solutions
of the flow equation which include infinitely many couplings; in the
LPA$^{\prime}$ case they are generic functions of the field. In order to find
such solutions one has typically to solve a differential equation
coming from the flow equation and integrate it numerically. More details
are given in section \ref{sub:Scaling-solutions}. 
In section \ref{sub:Some-simple-composite-operators} we discuss some
simple solutions of the composite operator flow equation.
In appendix \ref{sub:Critical-and-tri-critical-Ising-2d} we consider 
some numerical results obtained in the literature within the LPA$^{\prime}$ approximation 
for some statistical systems at criticality
and discuss them in connection to our framework.

\subsection{Scaling solutions \label{sub:Scaling-solutions}}

We consider a scalar field theory and limit our discussion to the so called LPA$^{\prime}$ truncation
where, we take into account up to two derivatives of the field and
a generic potential including the wave function renormalization:
\begin{eqnarray*}
\Gamma_{k}\left[\varphi\right] & = & \int\left[\frac{Z_{k}}{2}\partial_{\mu}\varphi\partial^{\mu}\varphi+U_{k}\left(Z_{k}^{1/2}\varphi\right)\right]\,.
\end{eqnarray*}
Let us denote $\tilde{\phi}=Z_{k}^{1/2}\tilde{\varphi}$. The flow
equation for the potential in dimensionless variables is given by
(throughout this work we consider the optimized cutoff \cite{Litim:2001up}) \cite{Ballhausen:2003gx,Codello:2012sc}:
\begin{eqnarray}
\partial_{t}\tilde{U}_{k} & = & -d\tilde{U}_{k}+\frac{d-2+\eta}{2}\tilde{\phi}\tilde{U}_{k}^{\prime}+c_{d}\frac{1-\frac{\eta}{d+2}}{1+\tilde{U}_{k}^{\prime\prime}}\label{eq:dimless-potential-flow-eq}\\
\eta & = & -\frac{\partial_{t}Z_{k}}{Z_{k}}=c_{d}\frac{\left(\tilde{U}_{k}^{\prime\prime\prime}\right)^{2}}{\left(1+\tilde{U}_{k}^{\prime\prime}\right)^{4}}\nonumber \, .
\end{eqnarray}
where $c_{d}^{-1}=\left(4\pi\right)^{d/2}\Gamma\left(d/2+1\right)$ and the field has been
set to its minimum in the equation for $\eta$.
The prime denotes a derivative with respect to the argument. 

As far as the results obtained with this truncation are concerned
the situation is the following: some of the critical exponents are already
in (relative) quantitative agreement with exact/best values available while others are less
precise. The anomalous dimension $\eta$ has usually a rather large error. 
This is a known shortcoming of the LPA$^{\prime}$ truncation which can
be overcome with more general ans{\"a}tze and/or employing more refined
truncation schemes as those developed in \cite{Blaizot:2005wd,Zambelli:2015ypo}.

It is worth to observe that the derivation of the flow equations (\ref{eq:flow_eq_composite_operator_epsilon})
and (\ref{eq:lambda_composite_operator_integrated}) is very similar
to the linearization of the flow equation itself and thus to the linearized
RG and the associated critical exponents. Let us spell out the relation
between the equation of eigenperturbation of the RG and composite
operators for the LPA$^{\prime}$ truncation. Let $\delta\tilde{U}_{k}=\left(k/k_{0}\right)^{-\theta}\delta v$
be the eigenperturbation, then the linearized form of the equation
reads:
\begin{eqnarray}
-\theta\delta\tilde{U}_{k} & = & -d\delta\tilde{U}_{k}+\frac{d-2+\eta}{2}\tilde{\phi}\delta\tilde{U}_{k}^{\prime}-c_{d}\frac{1}{1+\tilde{U}_{k}^{\prime\prime}}\delta\tilde{U}_{k}^{\prime\prime}\frac{1-\frac{\eta}{d+2}}{1+\tilde{U}_{k}^{\prime\prime}}\,.\label{eq:flow_eq_for_critical_exponents}
\end{eqnarray}
It is thus clear that under our approximations, i.e.~composite operators are parametrized as functions of the field,
the above flow equation
for $\delta\tilde{U}_{k}$ and equation (\ref{eq:lambda_composite_operator_integrated})
for $D_{i}$ are the same provided that we make the identification
$d-\theta_{i}=\lambda_{i}$. This relation can be extended to any
truncation with the following caveat. The linearized flow equation
for an eigenpertubation $\delta\tilde{U}_{k}$ coincides with the
one obtained for composite operators once we restrict the composite
operator source $\varepsilon$ to be constant, hence neglecting total
derivatives terms. Note that this implies by no means that such operators
are not important. As we shall see in the following, among these total
derivative operators we will find the descendant operators of the
field $\varphi$ as well as other interesting scaling operators. Moreover
in a non-perturbative setting it is difficult to argue whether or not an
operator gives a sizeable contribution.

\subsection{Some simple composite operators \label{sub:Some-simple-composite-operators}}

In this section we discuss some exact solutions to equation (\ref{eq:flow_eq_composite_operator_epsilon})
and the associated equivalent relations. Let us recall that we express
a renormalized composite operator $\left[O_{i}\right]$ via the following generic
parametrization $Z_{ij}O_{i}\left(Z_{k}^{1/2}\varphi\right)=Z_{ij}O_{j}\left(\phi\right)$,
a simple example was given in equation (\ref{eq:composite_operator_example_ItkZub}). 

To begin with we note that there are two very simple solutions to equation (\ref{eq:flow_eq_composite_operator_epsilon}). 
The first solution is the identity operator which
does not require any renormalization and as such its anomalous dimension
is zero. This solution corresponds to a constant solution of equation
(\ref{eq:flow_eq_for_critical_exponents}) with eigenvalue $\theta=d$
so that its anomalous dimension, given by $d-\theta$, is
simply zero. The second solution is the field $\phi$ itself. In this
case $Z_{ij}=1/Z_{k}^{-1/2}$ implying that the anomalous dimension
of the field operator is $-Z_{k}^{-1}\partial_{t}Z_{k}/2=\eta/2$
as expected. In terms of equation (\ref{eq:flow_eq_for_critical_exponents})
the solution is $\tilde{\phi}$; in this case the last term of (\ref{eq:flow_eq_for_critical_exponents})
vanishes and the eigenvalue problem is solved by $\theta=\left(d+2-\eta\right)/2$
and thus $\lambda=d-\theta=\left(d-2+\eta\right)/2$, this being the
full scaling dimension of the field operator. 

Along the same lines there are some further operators worth
commenting which are solutions of equation (\ref{eq:flow_eq_composite_operator_epsilon}).
The operator $O_{\Delta^{n}}\equiv Z_{O_{\Delta^{n}}} \Delta^{n}\left(Z_{k}^{1/2}\varphi\right)$, where $\Delta^n$ is the $n$th power of the Laplacian,
is a solution of equation (\ref{eq:flow_eq_composite_operator_epsilon})
with $Z_{O_{\Delta^{n}}}=1/Z_{k}^{1/2}$. Indeed once again we observe
that the RHS of (\ref{eq:flow_eq_composite_operator_epsilon}) vanishes
due to the fact that $O_{\Delta^{n}}$ is made of a single field $\varphi$.
The anomalous dimension of this operator is thus $\gamma_{O_{\Delta^{n}}}=\eta/2$
and the full scaling dimension is simply $2n+\left(d-2+\eta\right)/2$.
This suggests to identify the operators $O_{\Delta^{n}}$ with the descendants
of the field $\varphi$ of an hypothetical CFT describing the fixed point
(as we discuss in the appendix, section \ref{subsub:2Dmodels}, in $d=2$ these operators 
are secondary operators but are not all of them). 

Further consistent solutions for the flow equation for composite operators
can be constructed using derivative operators. In particular we want
to show that if $\left[O\right]$ is a renormalized composite operator, i.e.~an
exact solution of equation (\ref{eq:flow_eq_composite_operator_epsilon}),
then $\Delta\left[O\right]$ is also a renormalized composite operator. To see this
we have to check that $\Delta\left[O\right]$ is also an exact solution
of equation (\ref{eq:flow_eq_composite_operator_epsilon}). This can
be noted as follows: it is convenient to integrate by parts the source
dependent term of the EAA, namely: $\varepsilon\cdot\Delta\left[O\right]=\Delta\varepsilon\cdot\left[O\right]$.
If we call $\hat{\varepsilon}\equiv\Delta\varepsilon$ we notice that
the flow equation for $\Delta\left[O\right]$ is nothing but the flow
equation for $\left[O\right]$ written with the new source $\hat{\varepsilon}$.
Since the solution $\left[O\right]$ is valid for arbitrary sources and
therefore also for $\hat{\varepsilon}$, we have that $\Delta\left[O\right]$
is a solution of the flow equation as well.\footnote{Note that it is non trivial to build composite operators out of other
composite operators. A simple example is given by $\varphi$ which
is possibly the simplest ``composite'' operator. In particular,
given $\varphi$, simple products like $\varphi^{n}$ are not solutions
of equation (\ref{eq:flow_eq_composite_operator_epsilon}), whose RHS
induces new operators via the mixing.} 
In appedix \ref{sub:Critical-and-tri-critical-Ising-2d} we discuss these operators
in connection with the results known for some critical models in two and three dimensions.

Finally in the LPA$^{\prime}$ approximation there is always an eigendirection
associated to the derivative of the dimensionless potential, $\tilde{U}^{\prime}$,
with critical exponent $\theta=\left(d-2+\eta\right)/2$, see \cite{Hellwig:2015woa}
and \cite{Osborn:2011kw} for a general discussion including both the Wilsonian action and the EAA. 
Then the scaling dimension of $\left[U^{\prime}\right]$
is $\Delta_{\left[U^{\prime}\right]}=\left(d+2-\eta\right)/2$ whose
anomalous part reads $\gamma_{\left[U^{\prime}\right]}=-\eta/2$.
In our framework we can consider the ``equation of motion'' operator 
\begin{eqnarray*}
O_{\delta\Gamma_{k}} & = & \frac{\delta\Gamma_{k}}{\delta\varphi}\left[\varphi\right]=Z_{k}^{1/2}\frac{\delta\Gamma_{k}}{\delta\phi}\left[\phi\right]\,.
\end{eqnarray*}
In order to check that $O_{\delta\Gamma_{k}}$ is an exact solution
of equation (\ref{eq:flow_eq_composite_operator_epsilon}) we note
that the RHS of (\ref{eq:flow_eq_composite_operator_epsilon}) can
be found directly from the LHS using the known running of the EAA.
Indeed we observe the following:
\begin{eqnarray*}
\partial_{t}\left(\varepsilon\cdot\frac{\delta\Gamma_{k}}{\delta\varphi}\left[\varphi\right]\right) & = & \left(\varepsilon\cdot\partial_{t}\frac{\delta\Gamma_{k}}{\delta\varphi}\left[\varphi\right]\right)\\
 & = & \varepsilon\cdot\left(-\frac{1}{2}G_{k}\cdot\frac{\delta^{3}\Gamma_{k}}{\delta\varphi^{3}}\left[\varphi\right]\cdot G_{k}\cdot\partial_{t}R_{k}\right)\,.
\end{eqnarray*}
The last term of this expression is exactly the RHS of equation (\ref{eq:flow_eq_composite_operator_epsilon})
for the operator $O_{\delta\Gamma_{k}}$. This means that no other
operator mixes with $O_{\delta\Gamma_{k}}$, which is thus an exact
solution of the equation. In particular we note that $Z_{k}^{1/2}$
can be identified with the mixing matrix $Z_{ij}$ and that
the associated anomalous dimension is $\gamma_{O_{\delta\Gamma_{k}}}=-\eta/2$.
The scaling dimension of $O_{\delta\Gamma_{k}}$ is $\Delta_{O_{\delta\Gamma_{k}}}=\left(d+2-\eta\right)/2$.
Let us note that this operator is clearly redundant since the redefinition
$\varphi\rightarrow\varphi+\varepsilon$ of the field can eliminate
this term from the effective action: 
\begin{eqnarray*}
\Gamma_{k}\left[\varphi+\varepsilon\right] & = & \Gamma_{k}\left[\varphi\right]+\varepsilon\cdot\frac{\delta\Gamma_{k}\left[\varphi\right]}{\delta\varphi}\,.
\end{eqnarray*}
Being redundant this operator should not be considered in the spectrum
of observable scaling operators, see also \cite{Osborn:2011kw}.
In appendix \ref{sub:Critical-and-tri-critical-Ising-2d} we report some examples where
such an operator is identified within the LPA$^{\prime}$ truncation. However, the identification of this redundant operator
may not be so straightforward in other truncations.

\section{Conclusions and outlooks \label{sec:Conclusions-and-outlooks}}

In this work we have considered the dependence of the EAA on the floating
normalization point $\mu$ at which the boundary condition is imposed.
Particular attention has been paid to the renormalization of composite operators. 
In section \ref{sec:RG-flow-of-composite-operators}
we have described the general features of the flow equation for the EAA
generalized to include sources for composite operators. In section
\ref{sec:Floating-normalization-point-and-scaling} we have shown
that the EAA satisfies a sort of Callan-Symanzik equation which entails
the invariance under changes of the floating normalization point $\mu$.
This mechanism unveils how anomalous scaling shows up in the EAA formalism.
We have also shown how the scaling of composite operators is related
to critical exponents. 
Finally, in section \ref{sec:Scaling-solutions-and-composite-operators}
we have considered the local potential approximation in view
of our discussion and we have found some simple solutions
to the flow equation for composite operators. 
It turns out that one can systematically
identify a redundant operator present in the spectrum of eigenperturbations
(as already observed in \cite{Osborn:2011kw}) and it is possible to straightforwardly
extend the solutions of the equation to include a class of total derivatives
operators which we identify with the descendants of primary operators
of the fixed point theory. 

The invariance under changes of the floating normalization point $\mu$
makes explicit an intriguing link between standard quantum field theory and
the FRG approach. Indeed, when the scale $k$ is lowered to zero one
is left with the standard effective action which satisfies the Callan-Symanzik
equation. In the FRG scheme the Callan-Symanzik equation involves
infinitely many couplings whereas more common schemes involve just
a finite set of couplings (working with a renormalizable theory).
The latter possibility, being perturbation theory a particular solution of the
flow equation, can be recovered provided one solves the flow iteratively
as outlined in \cite{Litim:2001ky,Litim:2002xm,Codello:2013bra,Papenbrock:1994kf}. We feel that this link between a
Wilsonian type of RG and the Gell-Mann and Low formulation goes in the
direction outlined in \cite{Benettin:1976qz}.

We also remark that, in principle, our analysis tells that one is
not allowed to discard total derivative operators in the spectrum
of eigenpertubations. Depending on the cases these operators may or
may not give important corrections to the scaling dimensions of fixed
point scaling operators. However, in a non-perturbative setting, as
the FRG is, we think that one should be aware of these possibly important
contributions.

Finally it may be interesting to consider some particular operators
like the stress energy tensor (see \cite{Rosten:2014oja,Sonoda:2015pva} for a FRG perspective)
or non local operators like Wilson loops or Polyakov loops (see \cite{Pawlowski:2005xe}). 
This goes however beyond the scope of the present work. 
Possibly the flow equation
(\ref{eq:flow_eq_composite_operator_epsilon}) could be applied in
gauge theory to test the approximate restoration of BRS invariance
in the limit $k\rightarrow0$. In this case one needs to evaluate
the BRS composite operators by coupling them to a (Grassmannian odd)
source and following the flow down to $k\rightarrow0$.
Finally, given the flexible nature of the EAA formalism, we feel that the reasonings outlined
in this work may be applied to many areas of interest.


\appendix

\section{$\varepsilon^{2}$ terms \label{sec:eps2-terms}}

So far we have been studying the flow of the modified action $S+\varepsilon\cdot O$
up to first order. Multiple insertions of the composite operators
can be obtained by several functional differentiations with respect
to the source $\varepsilon$. Thus it is worth studying also the RG
flow of higher order terms in $\varepsilon$. Let us recall
\begin{eqnarray*}
\Gamma\left[\varphi,\varepsilon\right] & = & J\cdot\varphi-W\left[J,\varepsilon\right]\,,\;\varphi=\delta_{J}W\\
\delta_{\varepsilon}\Gamma\left[\varphi,\varepsilon\right] & = & -\delta_{\varepsilon}W\left[J,\varepsilon\right]\,.
\end{eqnarray*}
If we take a further functional derivative we obtain:
\begin{eqnarray*}
\delta_{\varepsilon}^{2}\Gamma\left[\varphi,\varepsilon\right] & = & -\frac{\delta^{2}W\left[J,\varepsilon\right]}{\delta\varepsilon^{2}}-\frac{\delta^{2}W\left[J,\varepsilon\right]}{\delta J\delta\varepsilon}\cdot\frac{\delta J}{\delta\varepsilon}\\
 & = & -\frac{\delta^{2}W\left[J,\varepsilon\right]}{\delta\varepsilon^{2}}-\left(\frac{\delta}{\delta\varepsilon}\frac{\delta W\left[J,\varepsilon\right]}{\delta J}\right)\cdot\frac{\delta J}{\delta\varepsilon}\\
 & = & -\frac{\delta^{2}W\left[J,\varepsilon\right]}{\delta\varepsilon^{2}}-\left(\frac{\delta}{\delta\varepsilon}\varphi\right)\cdot\frac{\delta J}{\delta\varepsilon}=-\frac{\delta^{2}W\left[J,\varepsilon\right]}{\delta\varepsilon^{2}}\,. 
\end{eqnarray*}
In the last line we used the fact that $\varphi=\delta_{J}W\left[J,\varepsilon\right]$
is a given function and thus it has no dependence on the source $\varepsilon$.
We see that crucial information regarding the insertion of two composite
operators can be obtained straightforwardly deriving twice with respect
to $\varepsilon$. Let us consider the flow equation: 
\begin{eqnarray}
\partial_{t}\left(\frac{\delta^{2}}{\delta\varepsilon^{2}}\Gamma_{k}\left[\varphi,\varepsilon\right]\right) 
 =\left[-\frac{1}{2}G_{k}\cdot\frac{\delta^{2}\Gamma_{k}^{\left(2\right)}}{\delta\varepsilon^{2}}\cdot G_{k}\cdot\partial_{t}R_{k}+G_{k}\cdot\frac{\delta\Gamma_{k}^{\left(2\right)}}{\delta\varepsilon}\cdot G_{k}\cdot\frac{\delta\Gamma_{k}^{\left(2\right)}}{\delta\varepsilon}\cdot G_{k}\cdot\partial_{t}R_{k}\right] . \label{eq:flow_eq_second_order_epsilon}
\end{eqnarray}
If we expand the EAA in terms of $\varepsilon$ we obtain an expression
of the following form:
\begin{eqnarray}
\Gamma_{k}\left[\varphi,\varepsilon\right] & = & \Gamma_{k}\left[\varphi\right]+\int_{x}\varepsilon\left(x\right)O_{k}\left(x\right)+\int_{x,y}\varepsilon\left(x\right)\varepsilon\left(y\right)B_{k}\left(x,y\right)+O\left(\varepsilon^{3}\right)\,.\label{eq:EAA_up_to_2nd_order_epsilon}
\end{eqnarray}
The flow equation (\ref{eq:flow_eq_composite_operator_epsilon}) gives
us the running of $O_{k}$ appearing at the first order in $\varepsilon$.
The term $B_{k}$ in (\ref{eq:EAA_up_to_2nd_order_epsilon}) can be
determined from equation (\ref{eq:flow_eq_second_order_epsilon}).
Let us note that if we set
\begin{eqnarray*}
\Gamma_{k}\left[\varphi,\varepsilon\right] & = & \Gamma_{k}\left[\varphi\right]+\int_{x}\varepsilon\left(x\right)O_{k}\left(x\right)+\int_{x}\varepsilon\left(x\right)^{2}B_{k}\left(x\right)+O\left(\varepsilon^{3}\right)
\end{eqnarray*}
the vertex in the first term of (\ref{eq:flow_eq_second_order_epsilon})
would amount just to a contact term, i.e.~a term proportional to
the Dirac delta. As such we could discard this term at separate points.
However, it is important to stress that most likely it is crucial to
keep $B_{k}\left(x,y\right)$ as a semilocal (as opposed to local) term. 
A simple example of this is given in \cite{Rosten:2010vm} where
the author considers the Wilsonian action keeping track of the source
$J$ conjugate to the field. In order to obtain the correct two-point
correlation function from two $J-$differentiations, it is crucial to keep
a semilocal term at order $J^{2}$.

Equation (\ref{eq:flow_eq_second_order_epsilon}) is potentially the
starting point to study in a nonperturbative setting the correlation
functions with two insertions of composite operators. In turn this
implies the possibility of studying the operator product expansion
coefficients along the lines of \cite{Hughes:1988cp,Keller:1991bz,Keller:1992by,Hollands:2011gf}. 
In this sense the study
of the $O\left(\varepsilon\right)$ terms is a first step in this direction.

\section{LPA and composite operators \label{sub:Critical-and-tri-critical-Ising-2d}}

In this appendix we consider the results of the LPA$^{\prime}$ truncation and
compare them with the exact results coming from conformal field theory
(CFT). In particular we shall consider the critical and tricritical
Ising model in two dimensions and the Ising model in three dimensions.
According to the discussion of sections \ref{sub:Fixed-point-and-anomalous-dim},
\ref{sub:Scaling-dimension-from-flow-eq} and \ref{sub:Scaling-solutions}
in the LPA$^{\prime}$ approximation for composite operators we are led
to identify the full dimension of the scaling operators $\lambda_{i}$
as $d-\theta_{i}$, where $\theta_{i}$ is a critical exponent. It
is thus straightforward to use the known results regarding critical
exponents to deduce anomalous dimensions of composite operators under
these approximations. We shall consider the results of reference \cite{Hellwig:2015woa}
whose methods allow to find several eigendirections in a systematic
manner, see also \cite{Litim:2003kf}.

\subsection{Critical and tricritical Ising models} \label{subsub:2Dmodels}

In two dimensions, by means of CFT techniques \cite{Belavin:1984vu}, it has been possible
to exactly compute the scaling dimension of the operators in the theory.
In particular the critical and tricritical Ising models correspond
to the minimal models having central charge $c=1-6/\left(m\left(m+1\right)\right)$
with $m=3$ and $m=4$. In the following we compare the exact results
with our approximations and comment the results. The correspondence
between composite operators of the Landau-Ginzburg Hamiltonian and
the scaling fields of the CFT is due to Zamolodchikov \cite{Zamolodchikov:1986db}.

Let us consider the Ising model. Table \ref{table:Ising2D} shows the first four
scaling operators (the identity operator is not shown) 
found using equations (\ref{eq:lambda_composite_operator_integrated}) and (\ref{eq:flow_eq_for_critical_exponents}).
As already anticipated the estimate for the anomalous dimension
of the field has a large error. This is a known feature of the LPA$^{\prime}$
truncation and better results can be obtained by employing a more
general kinetic term of the form $K\left(\varphi\right)\partial\varphi\partial\varphi$ \cite{Morris:1994jc}. 
{ \renewcommand{\arraystretch}{1.2}
\begin{table}
\begin{center}
\begin{tabular}{ | l | c | c | c | c | }
\hline
operator   & exact &  LPA$^{\prime}$ \\ \hline\hline
$\left[\varphi\right]\sim\phi_{2,2}$ & $\frac{1}{8}=0.125$ & $\frac{\eta}{2}=0.22$ \\  \hline
$\left[\varphi^{2}\right]\sim\phi_{1,3}$ & $1$ & $1.05$ \\ \hline
$\left[\varphi^{3}\right]$ &  & $1.78$ \\ \hline
$\left[\varphi^{4}\right]$ &  & $2.68$ \\ \hline
\end{tabular}
\end{center}
\caption{Scaling dimension in the critical Ising model.
The first column indicates the composite operator whose exact scaling dimension is reported in the second column. The 
third column lists the scaling dimensions obtained within the LPA$^{\prime}$ approximation using the critical exponents computed in
reference \cite{Hellwig:2015woa}.} \label{table:Ising2D}
\end{table}
}

Given that the anomalous dimension has such an error
one may expect that also the critical exponents, and thus the anomalous
dimension of composite operators, are not precise. 
Actually this depends
on which quantity we consider: certain quantities converge to relatively
precise values already in simple truncations while others need more
refined approximations. 
In the present
case we observe that the anomalous dimension of $\left[\varphi^{2}\right]$
is close to its correct value. The operator that we denoted $\left[\varphi^{3}\right]$
is simply the redundant operator $O_{\delta\Gamma_{k}}$ and as such
it does not appear among the physical scaling operators. 

Note that the arguments outlined in section \ref{sub:Some-simple-composite-operators}
allow us to easily identify some of the descendant operators associated to
$\left[\varphi\right]$ and $\left[\varphi^{2}\right]$. 
More precisely our arguments identify the secondary operators which are not quasi-primaries
(these are derivative operators of the type $L_{-1} \Phi$, where $L_n$ are the generators
of the Virasoro algebra and where we omitted the antiholomorphic generator). 
Other operators, like $L_{-2}\phi_{2,2}$,
should be present in the spectrum of eigenoperators and in principle should be seen.
Unfortunately, the other
operators present in the spectrum of equation (\ref{eq:flow_eq_for_critical_exponents})
have scaling dimensions which are not easily put in correspondence with CFT results.
As noted in \cite{Morris:1994jc} this may be due to the fact that higher dimension operators
correspond to operators having also many derivatives, which are not present in our truncation.
Ideally, solving the flow equation for composite operators one should find the spectrum 
of scaling dimensions known from CFT together with the associated degeneracy at
each level. Of course this is an incredibly hard task, but one can aim to obtain approximate results.

{ \renewcommand{\arraystretch}{1.2}
\begin{table}
\begin{center}
\begin{tabular}{ | l | c | c | c | c | }
\hline
operator   & exact &  LPA$^{\prime}$ \\ \hline\hline
$\left[\varphi \right] \sim\phi_{2,2}$ & $\frac{3}{40}=0.075$ & $\frac{\eta}{2}=0.156$ \\  \hline
$\left[\varphi^{2}\right]\sim\phi_{3,3}$ & $\frac{1}{5}=0.2$ & $0.33$ \\ \hline
$\left[\varphi^{3}\right]\sim\phi_{2,1}$ & $\frac{7}{8}=0.875$ & $0.84$ \\ \hline
$\left[\varphi^{4}\right]\sim\phi_{3,2}$ & $\frac{6}{5}=1.2$ & $1.32$ \\ \hline
$\left[\varphi^{5}\right]$ &  & $1.84$  \\ \hline
$\left[\varphi^{6}\right]\sim\phi_{3,1}$ & $3$ & $2.45$   \\ \hline
\end{tabular}
\end{center}
\caption{Scaling dimension in the tricritical Ising model.
The first column indicates the composite operator whose exact scaling dimension is reported in the second column. The 
third column lists the scaling dimension obtained within the LPA$^{\prime}$ approximation using the critical exponents computed in
reference \cite{Hellwig:2015woa}.} \label{table:TriIsing2D}
\end{table}
}
We shall now consider a similar analysis for the tricritical Ising
model. The exact values are compared with the results of the LPA$^{\prime}$ truncation
in table \ref{table:TriIsing2D}.
We observe that the anomalous dimension of $\left[\varphi \right]$ is by about a factor of two bigger
than the exact result, a more refined computation yields much better
predictions \cite{Morris:1994jc}. 
We observe that besides $\eta$ also the anomalous
dimension of $\left[\varphi^{2}\right]$ is rather poor, while those for $\left[\varphi^{3}\right]$
and $\left[\varphi^{4}\right]$ are closer to the exact values. Once
again better values can be found by considering a more refined truncation
which includes mixing with derivatives \cite{Morris:1994jc}. The operator
$\left[\varphi^{5}\right]$ can be identified with the redundant operator
$O_{\delta\Gamma_{k}}$. The discussion regarding descendant operators
that we did for the critical Ising model applies also in this case.

\subsection{Three dimensional Ising model \label{sub:Three-dimensional-Ising-model}}

The three dimensional Ising model is a paradigmatic application of the
renormalization group and has been studied via various truncations
in the FRG literature. These studies allowed a rather precise determination
of the anomalous dimension and the critical exponents \cite{Morris:1994ie,Morris:1994ki,Bonanno:2000yp,Canet:2003qd,Litim:2010tt}. 
In this section we use the results of reference \cite{Hellwig:2015woa} to
obtain the anomalous dimension of composite operators following the
discussion of section \ref{sub:Fixed-point-and-anomalous-dim}. The
results are shown in table \ref{table:Ising3D} where we compare the LPA$^{\prime}$ approximation
with the rigorous results found by conformal bootstrap techniques 
\cite{ElShowk:2012ht,El-Showk:2014dwa,Gliozzi:2013ysa,Gliozzi:2014jsa}\footnote{Interestingly a proof of conformal invariance of the Ising model has been given in \cite{Delamotte:2015aaa} using functional 
renormalization techniques.}.
{ \renewcommand{\arraystretch}{1.2}
\begin{table}[h]
\begin{center}
\begin{tabular}{ | l | c | c | c | c | }
\hline
operator & bootstrap & operator & LPA$^{\prime}$  \\ \hline\hline
$\sigma$ & $0.52$ & $\left[ \varphi\right]$ & $0.56$ \\  \hline
$\varepsilon$ & $1.41$ & $\left[\varphi^{2}\right]$ & $1.45$  \\ \hline
 &  & $\left[\varphi^{3}\right]$ & $2.44$  \\ \hline
$\varepsilon^{\prime}$ & $3.83$ & $\left[\varphi^{4}\right]$ & $3.53$\\ \hline
$\sigma^{\prime}$ & $4.05$ & $\left[\varphi^{5}\right]$ & $4.69$ \\ \hline
$\varepsilon^{\prime\prime}$ & $
\begin{array}{c} 
4.61\\
\approx7 
\end{array}$ & $\left[\varphi^{6}\right]$ & $5.9$   \\ \hline
\end{tabular}
\end{center}
\caption{Scaling dimensions in the three dimensional Ising model.
The first column indicates the (primary) operators whose scaling dimension is reported in the second column. The results come 
from the bootstrap approach \cite{ElShowk:2012ht,El-Showk:2014dwa,Gliozzi:2013ysa,Gliozzi:2014jsa}. 
(Only the first two digits are shown but more can be found in \cite{El-Showk:2014dwa,Gliozzi:2014jsa}). 
In the third column we indicate the composite operators whose scaling dimension is obtained from the LPA$^{\prime}$ approximation using the critical
exponents computed in reference \cite{Hellwig:2015woa}.} \label{table:Ising3D}
\end{table}
}

As in the other models that we have discussed, the anomalous dimension
of the field is poorly determined under our approximations. We note
that $\left[\varphi^{2}\right]$ is relatively close to the exact value
while for the other operators the results are not so precise. The
operator $\left[\varphi^{3}\right]$ can be identified
with the redundant operator $O_{\delta\Gamma_{k}}$. Being redundant
this operator must be discarded and indeed it finds no counterpart
in the part of table \ref{table:Ising3D} dedicated to the bootstrap approach. Furthermore
the result for $\left[\varphi^{6}\right]$ has a large error compared
to the bootstrap results \cite{El-Showk:2014dwa,Gliozzi:2014jsa}\footnote{
In the last row of table \ref{table:Ising3D} two values coming from the bootstrap approach are shown. 
The difference in the result depends on whether an operator disappears
or not from the spectrum \cite{Gliozzi:2014jsa}. 
This discrepancy should eventually disappear.
Our result is however rather ``wrong'' in both cases.}. 
Moreover the arguments of section \ref{sub:Some-simple-composite-operators} allow us 
to easily up-grade our solution to include the descendants of the fields in table \ref{table:Ising3D}.
However, given the difficulties encountered with the two dimensional Ising model, 
we feel that one should take the identifications of table \ref{table:Ising3D} with a grain of salt,
especially regarding $\left[\varphi^{6}\right]$. 

\section*{Acknowledgments}

We would like to thank Martin Reuter for illuminating discussions and suggestions
and Andreas Nink for carefully reading the paper and for discussions.

\end{document}